\documentclass[%
preprint,
 bibnotes,
 amsmath,
 amssymb,
 aps,
 prstab,
 floatfix,
 longbibliography
]{revtex4-1}
\usepackage{graphicx}% Include figure files
\usepackage{epsfig}
\usepackage{dcolumn}% Align table columns on decimal point
\usepackage{bm}% bold math
\usepackage{hyperref}% add hypertext capabilities
\usepackage{amsmath}
\usepackage{longtable}
\usepackage{mathtools}
\usepackage{siunitx}
\usepackage{appendix}
\usepackage{todonotes}
\newcommand{\E}{\mathcal{E}}
\newcommand{\F}{\mathcal{F}}
\newcommand{\TF}{\Tilde{\mathcal{F}}}

\begin{document}

\title{Thresholds for loss of Landau damping  in longitudinal plane}
% and decoherence
\author{Ivan Karpov\thanks{ivan.karpov@cern.ch}, Theodoros Argyropoulos, and Elena Shaposhnikova}
\affiliation{CERN, CH 1211 Geneva 23, Switzerland}

\date{\today}

\begin{abstract}

Landau damping mechanism plays a crucial role in providing single-bunch stability in LHC, High-Luminosity LHC, other existing as well as previous and future (like FCC) circular hadron accelerators. In this paper,
the thresholds for the loss of Landau damping (LLD) in the longitudinal plane are derived analytically using the Lebedev matrix equation (1968) 
and the concept of the emerged van Kampen modes (1983).
We have found that 
for the commonly-used particle distribution functions from a binomial family,
the LLD threshold vanishes in the presence of the constant inductive impedance $\mathrm{Im}Z/k$ 
above transition energy.  Thus, the effect of the cutoff frequency or the resonant frequency of a broad-band impedance on beam dynamics is studied in detail.
The findings are confirmed by direct numerical solutions of the Lebedev equation as well as using the Oide-Yokoya method (1990).
Moreover, the characteristics, which are important for beam operation, as the amplitude of residual oscillations and the damping time after a kick (or injection errors) are considered both above and below the threshold.
Dependence of the threshold on particle distribution in the longitudinal phase space is also analyzed, including some special cases with a non-zero threshold for Im$Z/k = {\it const}$.
All main results are confirmed by macro-particle simulations and consistent with available beam measurements in the LHC.

\end{abstract}
\maketitle
%%%%%%%%%%%%%%%%%%%%%%%%%%%%%%%%%%%%%%%%%%%%%%%%%%%%%%%%%%%%
\section{\label{sec:Introduction}Introduction}
\noindent
%%%%%%%%%%%%%%%%%%%%%%%%%%%%%%%%%%%%%%%%%%%%%%%%%%%%%%%%%%%%

Landau damping \cite{Landau} is one of the most efficient mechanisms of beam stabilization in both transverse and longitudinal planes of particle motion in hadron synchrotrons.
Landau damping is considered to be lost when the frequency of the coherent bunch oscillations   moves outside
the incoherent frequency band.

In the longitudinal plane, Landau damping of the coherent modes is provided by
synchrotron frequency spread coming from a non-linearity of the rf field, modified by a self-induced wake-field.
Synchrotron frequency spread can be increased by increasing the rf-bucket filling factor or longitudinal emittance and by using a higher-harmonic rf system. All these (``passive") methods are applied in CERN synchrotrons together with some ``active" damping systems of beam instabilities (feedback systems).
 
Undamped coherent beam motion observed at Tevatron \cite{Tevatron}, RHIC \cite{RHIC}, SPS~\cite{SPS} and LHC \cite{LHC} can be attributed to loss of Landau damping (LLD).
In LHC, in the absence of a longitudinal wide-band feedback system and a higher harmonic rf system, single bunch stability is guaranteed by controlled longitudinal emittance blowup during the acceleration cycle.   
However, it was noticed from the start of the LHC operation that the beams with the nominal parameters are at the limit of stability at different moments of the acceleration cycle due to LLD, with thresholds about four times lower than expected \cite{MCBI2019}.
An agreement of first macro-particle simulations, based on the existing impedance model, with the measurements \cite{JMPhD}  suggested that the discrepancy cannot be explained by a large missing impedance. 
The initial choice of the longitudinal beam parameters in the LHC~\cite{LHCDR} was based on the criterion derived from the Sacherer stability diagrams~\cite{Sacherer2} (see Ref.~\cite{NG} for more details). While in the longitudinal plane these diagrams are justified   only for a coasting beam \cite{Keil} or a very low-frequency impedance \cite{MCBI2019}, the Sacherer criterion is still widely used for the  analytical threshold evaluation,  also often assuming a constant impedance $\text{Im}Z/k$, with $k$ being a harmonic of the revolution frequency.

A general way to study beam stability is to use the Vlasov equation linearized for a small perturbation of a stationary particle distribution function. 
The first self-consistent system of equations suitable for the eigenvalue analysis of 
longitudinal beam stability was proposed by A.~N.~Lebedev in 1968 \cite{Lebedev1968}.
Converting it to an integral equation, he has also proven that a space-charge impedance cannot cause instability of a bunched beam. Nevertheless,
as was shown later, a reactive impedance may lead to LLD.

In the early 70's, F.~Sacherer derived another matrix equation, from which then the stability diagrams  
were obtained using the following assumptions:
(i) the  rf field is linear;
(ii) the actual \emph{self-force} (from beam interaction with its environment) 
depends only on the location of the bunch center-of-mass (so-called ``synthetic kernel" approach describing the rigid-bunch motion); (iii) the synchrotron frequency spread is neglected in the calculation of coherent frequency shift with intensity.
The LLD threshold is reached when the coherent mode crosses the boundary of the stability diagram.
The LLD threshold, similar to the Sacherer analytical criterion, was also obtained from the integral form of the Lebedev equation \cite{Lebedev1968} in Ref.~\cite{LLD_BI_EPAC90} for a constant $\text{Im}Z/k$. However, as was shown recently in Ref.~\cite{MCBI2019}, the Sacherer stability diagrams can be obtained from the Lebedev equation using the low-frequency approximation.

Another approach was suggested by A. Hofmann and F. Pedersen~\cite{Hofmann1979} and used more recently, for example, for analysis of Landau damping in single \cite{Oliver1} and multi-harmonic rf systems \cite {Oliver2}. It is based on direct comparison of the coherent dipole frequency of the rigid-bunch oscillations with the maximum incoherent frequency inside the bunch, both calculated 
for a constant reactive impedance $\text{Im}Z/k$ and an elliptic particle distribution in longitudinal phase space (corresponding to the so-called ``parabolic" bunches). Contrary to the Sacherer stability diagrams, here the synchrotron frequency spread is taken into account in the calculation of coherent frequencies.
For the same conditions, the LLD threshold obtained by this method is only about 20-30\% higher than the one from the stability diagram \cite{Sacherer2}. 
However, comparison with measurements in the LHC indicates that both these  approaches may significantly underestimate the real threshold \cite{MCBI2019}.

There are several ways to find the accurate solutions of the linearized Vlasov equation without neglecting the synchrotron frequency spread \cite{Lebedev1968}, \cite{Besnier}, \cite{YHChin1983}, \cite{Oide_Yokoya1990}. In 1983, Y.~H.~Chin, K.~Satoh, and K.~Yokoya introduced the concept of van Kampen modes~\cite{vKampen1,vKampen2}
for a description of bunch oscillations in the longitudinal plane~\cite{YHChin1983}. Landau damping at low intensities was explained as decoherence of the van Kampen modes,
while the LLD threshold is reached, once a van Kampen mode emerges from the band of the incoherent synchrotron frequencies. In 2010, A.~Burov \cite{burov2012van} used this concept together with an eigenvalue approach,
suggested for analysis of single-bunch instabilities by K.~Oide and K.~Yokoya \cite{Oide_Yokoya1990}, to evaluate the LLD thresholds for different impedance models and rf potentials, also taking into account a potential well distortion.
In particular, he found  that the LLD threshold calculated for the constant $\text{Im}Z/k$ is more than a factor of three lower than the one obtained from the Sacherer stability diagrams~\cite{Burov2}.

Even though in all previous studies a finite LLD threshold was obtained for a constant Im$Z/k$, we will show below that in most cases this threshold is zero unless some cutoff frequency is introduced (which effectively is also the case in all numerical calculations or particle simulations).
The thresholds and the analytical criterion for the LLD are obtained by applying,
for the first time, the concept of van Kampen modes to solve 
the Lebedev equation in its original matrix form~\cite{Lebedev1968}.
These results are compared with those obtained using the Oide-Yokoya method, macro-particle simulations, and the beam measurements in the LHC.

The paper is organized in the following way. After this {\bf Introduction}, the main equations of longitudinal motion and solutions for stationary particle distribution with intensity effects included, are presented in {\bf Section~\ref{sec:main_equations}}.  For completeness, in \textbf{Section~\ref{sec:perturbation}}, both the Lebedev matrix equation and the Oide-Yokoya equation are derived step-by-step from the linearized Vlasov equation using the perturbation formalism. 
{\bf Section~\ref{sec:Thesholds}} starts with a brief reminder of the formalism of emerged van Kampen modes.
Then, we derive an analytic expression for the LLD threshold in the presence of the constant inductive impedance using the Lebedev equation. It is compared with the LLD thresholds determined numerically solving the two matrix equations 
with examples for the LHC parameters. 
The dependence of the LLD threshold on the cutoff frequency of the constant inductive impedance is studied first.
A similar analysis is also performed for a broad-band resonator model and the results are compared with macro-particle simulations using CERN Beam Longitudinal Dynamics code BLonD~\cite{BLOND2020}. The influence of particle distribution on the threshold value is also investigated here.
In {\bf Section~\ref{sec:Impact_on_beam}}, the meaning of the LLD threshold for an accelerator operation is analyzed in detail by considering a bunch response to the most common,  rigid-dipole, perturbation (a kick due to a phase/energy error or noise excitation). Using the van Kampen modes as a basis for an expansion of the rigid-dipole mode, the impact of this perturbation on the beam is evaluated semi-analytically and then compared with macro-particle simulations. In particular, it is demonstrated that
the damping time of a perturbation below the LLD threshold and the amplitude of the residual bunch oscillations above this threshold are the most important characteristics of beam behavior. 
Moreover, the obtained results are also compared with available beam measurements in the LHC.
% which allow understanding the consequences of LLD for both existing and future particle accelerators.
The LLD thresholds for some interesting cases from a parameter space, which was not covered in the previous sections,  are  briefly discussed in {\bf Section~\ref{sec:discussions}}. This includes the reactive impedance with the opposite sign (capacitive above transition or inductive below)  and a special class of particle distribution functions, both leading to a finite LLD threshold even for a constant $\text{Im}Z/k$ impedance.
The main conclusions of this work are presented in {\bf Section~\ref{sec:conclusions}}.
In the end, {\bf Appendixes~\ref{annex:A}~and~\ref{annex:B}} contain some important derivations.
%The role of the realistic impedance model is also emphasised.

%Here we also analyse the impact of the rigid-dipole kick on the beam. By using a mode expansion, we predict analytically a bunch offset evolution after the kick. We demonstrate an excellent agreement with macro-particle simulations using the BLonD code \cite{BLOND2020}, but also point out limitations due to numerical noise. Below the LLD threshold, the damping time of perturbation can calculated, while above threshold the amplitude of residual oscillations  can be evaluated accurately.

%%%%%%%%%%%%%%%%%%%%%%%%%%%%%%%%%%%%%%%%
%%%%%%%%%%%%%%%%%%%%%%%%%%%%%%%%%%%%%%%%
\section{\label{sec:main_equations} Longitudinal motion with intensity effects }
%%%%%%%%%%%%%%%%%%%%%%%%%%%%%%%%%%%%%%%%
%%%%%%%%%%%%%%%%%%%%%%%%%%%%%%%%%%%%%%%%
%%%%%%%%%%%%%%%%%%%%%%%%%%%%%%%%%%%%%%%%%%%%%%%%%%%%%%%%%
% \subsection{Main equations and definitions}
%%%%%%%%%%%%%%%%%%%%%%%%%%%%%%%%%%%%%%%%%%%%%%%%%%%%%%%%%
We start with a stationary situation, while time-dependent perturbations are treated in the following section. We chose the coordinate system with respect to the synchronous particle with the energy $E_0$ and the phase $\phi_{s0}$ that corresponds to the case of a single rf system without taking into account intensity effects. Then, the main equations of longitudinal particle motion can be written using conjugate variables \{$\Delta E/(h\omega_0)$, $\phi$\},
\begin{flalign}
    \frac{d\phi}{dt}= \dot{\phi} &= \frac{h^2\omega^2_0\eta}{\beta^2 E_0}\left(\frac{\Delta E}{h\omega_0} \right)  \label{eq:first_eq_motion},\\
    \frac{d}{dt}\left(\frac{\Delta E}{h\omega_0} \right) &= \frac{1}{2\pi h}\left[qV_t(\phi) - \delta E_0\right]. \label{eq:second_eq_motion}
\end{flalign}
where $\Delta E$ and $\phi$ are respectively the energy and phase deviations of the particle, $\omega_0 = 2\pi f_0$, $f_0$ is the revolution frequency, $\beta$ is the particle velocity normalized by the speed of light, $h$ is the harmonic number, $q$ is the electrical charge of the particle, $\delta E_0$ is the energy gain per turn of the synchronous particle, $\eta = 1/\gamma^2_\mathrm{tr} - 1/\gamma^2$ is the slip factor, $\gamma$ is the relativistic Lorentz factor, and $\gamma_\mathrm{tr}$ is the Lorentz factor at transition energy. These equations can be found in standard textbooks (e.g.~\cite{Chao}).

The present work focuses on the case of a single bunch in a single rf system, while the derivations can be easily adapted to any combination of rf waves.
In the presence of intensity effects, the total voltage $V_t(\phi) = V_\mathrm{rf}(\phi) + V_\mathrm{ind}(\phi)$, in addition to the rf voltage $V_\mathrm{rf}$, contains contributions from the beam-induced fields described as the beam-induced voltage $V_\mathrm{ind}$. 
Then, $V_\mathrm{rf}(\phi) = V_0\sin(\phi_{s0} + \phi)$, with $V_0$ being the rf voltage amplitude, and the synchronous phase $\phi_{s0} = \arcsin{(\delta E_0/qV_0)}$ for $\eta<0$ or $\phi_{s0} =\pi-\arcsin{(\delta E_0/qV_0)}$ for $\eta>0$. The stationary beam-induced voltage can be written in the form (see Appendix~\ref{annex:A}),
\begin{equation}
    \label{eq:induced_voltage}
    V_\mathrm{ind}(\phi) =  \sum_{k=-\infty}^{\infty} V_k e^{i \frac{k}{h} \phi} = - q N_p \, h \, \omega_0 \sum_{k=-\infty}^{\infty} Z_k \lambda_k  e^{i \frac{k}{h} \phi},
\end{equation}
where $N_p$ is the total number of particles in the bunch, $Z_k = Z(k \omega_0)$ is the longitudinal impedance at frequency $k \omega_0$, and $\lambda_k$ is the Fourier harmonic of the normalized line density, $\lambda(\phi) = (d N/d\phi)/N_p$,
\begin{equation}
    \label{eq:line_density_harm}
    \lambda_k = \frac{1}{2\pi h} \int_{-\pi h}^{\pi h}d\phi \; \lambda(\phi) e^{-i \frac{k}{h} \phi}.
\end{equation}
The line density is defined by the distribution function $\F(\phi,\dot{\phi})$, i.e. $\lambda(\phi) = \int_{-\infty}^{\infty}d\dot{\phi}\, \F(\phi,\dot{\phi})$, and the following normalization is imposed
\begin{equation}
    \label{eq:normalization}
    \int_{-\pi h}^{\pi h}d\phi\, \lambda(\phi) = \int_{-\pi h}^{\pi h}d\phi \int_{-\infty}^{\infty}d\dot{\phi}\, \F(\phi,\dot{\phi}) = 1.
\end{equation}

It is convenient for further analysis to use another set of variables ($\mathcal{E}$, $\psi$),  which correspond respectively to the energy and phase of the synchrotron oscillations,
\begin{align}
    \mathcal{E} &= \frac{{\dot{\phi}}^2}{2\omega^2_{s0}} + U_t(\phi), \label{eq:energy} \\ 
    \psi &= \text{sgn}(\eta\Delta E)\frac{\omega_s(\mathcal{E})}{\sqrt{2}\omega_{s0}} \int_{\phi_\mathrm{max}}^\phi \frac{d\phi^\prime}{\sqrt{\mathcal{E} -U_t\left(\phi^\prime\right)}}. 
\label{eq:phase}
\end{align}
Here $\omega_{s0} = 2\pi f_{s0}$ is the angular frequency of small-amplitude synchrotron oscillations in a bare rf potential with 
\begin{equation}
    \label{eq:synchrotron_tune_0}
    \omega_{s0}^2 = -\frac{h \, \omega_0^2 \eta q V_0\cos{\phi_{s0}}}{2\pi\beta^2 E_0}.
\end{equation}
The total potential can be obtained from the total voltage $V_t$
\begin{equation}
    \label{eq:potential}
    U_t(\phi) = \frac{1}{V_0\cos{\phi_{s0}}}\int^\phi_{\Delta \phi_{s}} d\phi^\prime \, \left[V_t(\phi^\prime) - V_0\sin{\phi_{s0}} \right]
\end{equation}
with the synchronous phase shift due to intensity effects $\Delta \phi_s$ that satisfies the relation $V_0 \sin{\phi_{s0}} = V_0\sin(\phi_{s0} + \Delta \phi_s) + V_\mathrm{ind}(\Delta \phi_s)$. 
The dependence of the synchrotron frequency on the energy of synchrotron oscillations $\omega_s(\mathcal{E}) =2\pi/T_s(\mathcal{E})$ in Eq.~(\ref{eq:phase}) can be found from the period of oscillations 
\begin{equation}
    \label{eq:synchrotron_freq}
    T_s(\mathcal{E})= \frac{\sqrt{2}}{\omega_{s0}} \int_{\phi_{\min}(\E)}^{\phi_{\max}(\E)}\frac{d\phi^\prime}{ \sqrt{\E -U_t\left(\phi^\prime\right)} }, 
\end{equation}
where $\phi_{\min}(\E)$ and $\phi_{\max}(\E)$ are the minimum and maximum phases of the particle with the energy of synchrotron oscillation $\mathcal{E}$ which satisfy the relation $\mathcal{E}  = U_t(\phi)$.

The stationary particle distribution function $\F$ is
a function of only the energy of synchrotron oscillations $\E$
with the line density  
\begin{equation}
    \label{eq:norm_line_density}
    \lambda(\phi) = 2\omega_{s0} \int_{U_t(\phi)}^{\E_{\max}} d\E \frac{\F(\E)}{\sqrt{2\left[\E - U_t(\phi) \right]}}.
\end{equation}
It depends on the total potential that in calculations can be found using an iterative procedure, similar to the one used in Ref.~\cite{burov2012van} (see Appendix~\ref{annex:B}). Below we will mostly consider distributions belonging to a binomial family, which covers a wide range of realistic bunch shapes, from flat ($\mu = -1/2$) to  Gaussian ($\mu \to \infty $),
\begin{equation}
    \label{eq:binomial_distribution}
    \F(\E) = \frac{1}{2\pi \omega_{s0} A_N}\left(1 - \frac{\E}{\E_\mathrm{max}}\right)^\mu = \frac{g\left(\E\right)}{2\pi \omega_{s0} A_N},
\end{equation}
where the normalization factor
\begin{equation}
    \label{eq:normalization_factor}
    A_N = \omega_{s0}\int_0^{\E_{\max}}d\E \,\frac{  g\left(\E\right)}{\omega_s(\E)}.
\end{equation}
In this case, the integration in Eq.~(\ref{eq:norm_line_density}) can be performed analytically yielding
\begin{equation}
    \label{eq:line_density_binom}
    \lambda(\phi) = \frac{\sqrt{ \E_{\max}} \Gamma(\mu+1)}{\sqrt{2\pi}A_N\Gamma(\mu+3/2)} \left[1-\frac{U_t(\phi)}{\E_{\max}}\right]^{\mu+1/2},
\end{equation}
where $\Gamma$ is the gamma function.

For $\mu \to \infty$ a bunch has a Gaussian line density and the corresponding bunch length $\tau_{4\sigma}$ is typically defined as four times the Root-Mean-Square bunch length $\sigma$, i.e. $\tau_{4\sigma} = 4 \sigma$. The bunch length $\tau_{4\sigma}$ can be easily calculated from the Full-Width Half-Maximum (FWHM) bunch length $\tau_\mathrm{FWHM}$:
\begin{equation}
   \label{eq:bunch_length}
   \tau_{4\sigma} = \tau_\mathrm{FWHM} \sqrt{2/\ln 2}.
\end{equation}
In practice, the proton bunches are usually not Gaussian and have a finite full length $\tau_\mathrm{full}$ defined as
\begin{equation}
    \label{eq:full_bunch_length}
    \tau_\mathrm{full} = \left[\phi_{\max}(\E_{\max}) - \phi_{\min}(\E_{\max}) \right]/\omega_\mathrm{rf}.
\end{equation}
However, we will also use definition~(\ref{eq:bunch_length}) below as it has important features related to the LLD threshold (see Sec.~\ref{sec:Thesholds}).

%%%%%%%%%%%%%%%%%%%%%%%%%%%%%%%%%%%%%%%%%%%%%%%%%%%%%%%%%
\section{\label{sec:perturbation} Perturbation formalism}
%%%%%%%%%%%%%%%%%%%%%%%%%%%%%%%%%%%%%%%%%%%%%%%%%%%%%%%%%

%%%%%%%%%%%%%%%%%%%%%%%%%%%%%%%%%%%%%%%%%%%%%%%%%%%%%%%%%
% \subsection{The Vlasov equation}
%%%%%%%%%%%%%%%%%%%%%%%%%%%%%%%%%%%%%%%%%%%%%%%%%%%%%%%%%
For analysis of beam stability, 
we need to consider the perturbations $\Tilde{\F}$, $\Tilde{\lambda}$ and $\Tilde{V}_\mathrm{ind}$, respectively, to the equilibrium distribution function $\F$, line density $\lambda$, and induced voltage $V_\mathrm{ind}$, introduced in the previous section. If these deviations grow with time, the beam is unstable.

The linearized Vlasov equation in ($\E,\psi$) variables is
\begin{equation}
    \label{eq:vlasov_equation}
    \frac{\partial \Tilde{\F}}{\partial t} + \frac{d\E}{dt} \frac{d \F}{d\E} + \frac{d\psi}{dt} \frac{\partial \Tilde{\F}}{\partial \psi} = 0,
\end{equation}
where by definition $d\psi/dt =\omega_s(\E)$, while $d\E/dt$ can be found from a combination of the first and second equations of particle motion in the presence of perturbation
\begin{equation}
    \label{eq:second_eq_motion_perturbed}
    \frac{d\dot{\phi}}{dt} + \frac{\omega^2_{s0}}{V_0 \cos\phi_{s0}}
    \left[V_t(\phi) - V_0\sin{\phi_{s0}} \right] = - \frac{\omega^2_{s0}}{V_0 \cos\phi_{s0}} \Tilde{V}_\mathrm{ind}(\phi,t) .
\end{equation}
Multiplying both sides of Eq.~(\ref{eq:second_eq_motion_perturbed}) by $\dot{\phi}$
and taking into account Eq.~(\ref{eq:energy}) one obtains
%Taking time derivative of the relation $\dot{\phi} = \pm \sqrt{2}\omega_{s0}\sqrt{\E-U}$ %from Eq.~(\ref{eq:energy}) and combining it with %Eq.~(\ref{eq:second_eq_motion_perturbed}) one obtains
\begin{equation}
    \label{eq:energy_derivative}
    \frac{d\E}{dt} = -\frac{d\phi}{dt}\frac{\Tilde{V}_\mathrm{ind}(\phi,t)}{V_0 \cos\phi_{s0}}  = -\omega_s(\E)\frac{\partial \Tilde{U}_\mathrm{ind}(\phi,t)}{\partial \psi},
\end{equation}
where the perturbed induced potential $\Tilde{U}_\mathrm{ind}(\phi,t)$ is defined similarly to Eq.~(\ref{eq:potential})
\begin{equation}
    \label{eq:induced_potential}
    \Tilde{U}_\mathrm{ind} (\phi) = \frac{1}{V_0\cos{\phi_{s0}}}\int^\phi_{\Delta \phi_{s}}d\phi^\prime \Tilde{V}_\mathrm{ind}(\phi^\prime).
\end{equation}
Finally, the linearized Vlasov equation can be written in the form
\begin{equation}
    \label{eq:vlasov_equation_new}
    \left[\frac{\partial }{\partial t} + \omega_s \frac{\partial }{\partial \psi}\right] \Tilde{\F} = \omega_s \frac{\partial \Tilde{U}_\mathrm{ind}}{\partial \psi} \frac{d \F}{d\E}.
\end{equation}

Several approaches allow finding general solutions of this equation without neglecting the synchrotron frequency spread.
% and therefore to analyse Landau damping.
In the following subsection we derive, the equation that was proposed by A.~N.~Lebedev in 1968~\cite{Lebedev1968}, but since then it 
was never used in its original matrix form for evaluation of LLD for the single-bunch case.
%%%%%%%%%%%%%%%%%%%%%%%%%%%%%%%%%%%%%%%%%%%%%%%%%%%%%%%%%%%
\subsection{Lebedev equation}
%%%%%%%%%%%%%%%%%%%%%%%%%%%%%%%%%%%%%%%%%%%%%%%%%%%%%%%%%%%
Here we derive the matrix equation first obtained in Ref.~\cite{Lebedev1968}. The detailed derivation with slightly different variables can be also found in~\cite{Shaposhnikova1994}. 
The solution of Eq.~(\ref{eq:vlasov_equation_new}) must be periodic in $\psi$, and can be represented as the sum of the harmonics $e^{ i m \psi}$ ($m\neq 0$). 
Presenting the perturbations 
for a given $\Omega$ as $\Tilde{\F} (\E,\psi,t) = \Tilde{\F} (\E,\psi,\Omega) e^{-i\Omega t}$ and $\Tilde{U}_\mathrm{ind} (\E,\psi,t) = \Tilde{U}_\mathrm{ind} (\E,\psi,\Omega) e^{-i\Omega t}$,
we have
\begin{equation}
    \Tilde{\F}(\E,\psi,\Omega) = \sum_{m = -\infty}^{\infty} \Tilde{\F}_m (\E,\Omega) e^{i m \psi}, \;\;
    \Tilde{U}_\mathrm{ind} (\E,\psi,\Omega) = \sum_{m = -\infty}^{\infty} \Tilde{U}_{\mathrm{ind},m} (\E,\Omega) e^{i m \psi}, 
\end{equation}
with 
\begin{equation}
    \label{eq:harmonics_m}
    X_m (\E,\Omega) =\frac{1}{2\pi} \int_{-\pi}^{\pi}d\psi \, X(\E,\psi,\Omega) e^{-i m\psi}.
\end{equation}
Then the solution of Eq.~(\ref{eq:vlasov_equation_new}) is 
\begin{equation}
    \label{eq:solution}
    \Tilde{\F} (\E,\psi,\Omega) = - \omega_s (\E) \frac{d \F}{d\E} \sum_{m = -\infty}^{\infty}  \frac{m \Tilde{U}_{\mathrm{ind},m}(\E,\Omega) }{\Omega - m \omega_s(\E)}e^{i m\psi}.
\end{equation}
The perturbed induced voltage is related to the perturbed line density as (see Appendix~\ref{annex:A})
\begin{equation}
    \label{eq:ind_vs_line}
    \Tilde{V}_k(\Omega) = - q N_p \, h\, \omega_0 Z_k(\Omega) \Tilde{\lambda}_k (\Omega),
\end{equation}
where $Z_k(\Omega) = Z (k\omega_0 + \Omega)$. Then, using Eq.~(\ref{eq:induced_potential}), the perturbed induced potential can be presented in the form 
\begin{equation}
    \label{eq:ind_potential_perturb}
    \Tilde{U}_\mathrm{ind}(\E,\psi,\Omega) = -i\zeta\sum_{k=-\infty}^{\infty}\frac{Z_k(\Omega)/k}{\text{Im}Z/k} \Tilde{\lambda}_k(\Omega) \left( e^{i \frac{k}{h} \phi} - e^{i \frac{k}{h}\Delta \phi_{s}} \right),
\end{equation}
and its Fourier harmonics $\Tilde{U}_{\mathrm{ind},m}$, defined by Eq.~(\ref{eq:harmonics_m}), can be obtained as
\begin{equation}
    \label{eq:potential_hamonics_m}
    \Tilde{U}_{\mathrm{ind},m}(\E,\Omega) = -i\zeta \sum_{k=-\infty}^{\infty}\frac{Z_k(\Omega)/k}{\text{Im}Z/k} \Tilde{\lambda}_k(\Omega) I_{mk}(\E).
\end{equation}
Here we introduce the dimensionless ``intensity" parameter  $\zeta$,
\begin{equation}
    \label{eq:xi}
     \zeta = -\frac{ q N_p \, h^2 \, \omega_0 \, \text{Im} Z/k }{V_0 \cos\phi_{s0}}.
\end{equation}
Some constant value $\text{Im}Z/k$, used for the normalization of the impedance, is well defined for the case of constant inductive impedance considered in the present work, while it can be arbitrarily chosen for other impedance models.

The function $I_{mk}(\E)$, introduced for the first time in~\cite{Lebedev1968}, plays an important role 
\begin{equation}
    \label{eq:Imk}
    I_{m k} (\E) = \frac{1}{2\pi} \int_{-\pi}^{\pi} d\psi \; e^{i\frac{k}{h}\phi(\E,\psi) - i m \psi} = \frac{1}{\pi} \int_{0}^{\pi} d\psi \; e^{i\frac{k}{h}\phi(\E,\psi)} \cos m \psi.
\end{equation}
The last expression 
comes from the fact that $\phi(\E,-\psi) = \phi(\E,\psi)$. Note, that $I_{mk}$ depends on $\zeta$ because $\phi(\E,\psi)$ is modified via potential well distortion. It is easy to see also the following properties
\[
I_{-mk} = I_{mk},\; \text{and} \; I_{m-k} = I^*_{mk}.
\]
In the case of symmetric potential well, we also have $I^*_{mk} = (-1)^m I_{mk}$. 
An alternative way to evaluate $I_{mk}$ 
is to perform  in Eq.~(\ref{eq:Imk}) integration by parts
\begin{equation}
    \label{Imk_alternative}
    I_{mk}(\E) = \frac{i k}{h \pi m} \int_{\phi_{\min}(\E)}^{\phi_{\max}(\E)}d\phi\; e^{i\frac{k}{h}\phi} \, \sin[m\psi(\E,\phi)],
\end{equation}
where the phase of synchrotron oscillations is given by Eq.~(\ref{eq:phase}).

For perturbation of the line density at frequency $\Omega$
\begin{equation}
    \label{eq:line_density_perturb}
    \Tilde{\lambda}(\phi,\Omega) = \sum_{k=-\infty}^{\infty} \Tilde{\lambda}_k (\Omega) e^{i \frac{k}{h} \phi},
\end{equation} 
the harmonics $\Tilde{\lambda}_k(\Omega)$ are related to $\Tilde{\F}(\E,\psi,\Omega)$ as
\begin{equation}
    \label{eq:line_density_pert_harm}
    \Tilde{\lambda}_k (\Omega) = \frac{1}{2\pi h} \int_{-\pi h}^{\pi h}d\phi \, \Tilde{\lambda}(\phi) e^{-i \frac{k}{h} \phi} =  \frac{\omega_{s0}^2}{2\pi h} \int_{-\pi}^{\pi} d\psi \int_0^{\E_{\max}} d\E \; \frac{\Tilde{\F}(\E,\psi,\Omega)}{\omega_{s}(\E)} \, e^{-i\frac{k}{h}\phi(\E,\psi)}.
\end{equation}
Here, the transformation of variables  $d\phi d \dot{\phi} = \omega_{s0}^2 \, d\psi \,d\E/\omega_s(\E)$ was used.

At the next step, inserting Eq.~(\ref{eq:solution}) into Eq.~(\ref{eq:line_density_pert_harm}),  an infinite system of equations for harmonics of the line density perturbation can be obtained
\begin{equation}
    \label{eq:matrix_lebedev}
    \Tilde{\lambda}_p(\Omega) =- \frac{\zeta}{h} \sum_{k = -\infty}^{\infty} G_{p k} (\Omega) \, \frac{Z_k(\Omega)/k}{\text{Im}Z/k} \, \Tilde{\lambda}_k(\Omega),
\end{equation}
where the beam transfer matrices~\cite{Shaposhnikova1994} are defined as
\begin{align}
    &G_{p k} (\Omega) =  - i \, \omega_{s0}^2 \sum_{m=-\infty}^{\infty} m \int_0^{\E_{\max}}\! d\E \frac{d\F(\E)}{d\E} \frac{I_{m k}(\E) I^*_{m p}(\E)}{\Omega - m \omega_s(\E)} \nonumber \\
    &= - 2 i   \, \omega_{s0}^2 \sum_{m=1}^{\infty}  \int_0^{\E_{\max}} \!d\E \frac{d\F(\E)}{d\E} \frac{I_{m k}(\E) I^*_{m p}(\E) \omega_s(\E)}{\Omega^2/m^2 - \omega^2_s(\E)}. \label{eq:BTM}
\end{align}
After substitution of $\F$ from Eq.~(\ref{eq:binomial_distribution}) the element $G_{pk}$ has a form 
\begin{equation}
    \label{eq:BTM_new}
    G_{p k} (\Omega) = -  i \, \frac{\omega_{s0}}{\pi A_N} \sum_{m=1}^{\infty} \int_0^{\E_{\max}} d\E \frac{d g(\E)}{d\E} \frac{I_{m k}(\E) I^*_{m p}(\E) \omega_s(\E)}{\Omega^2/m^2 - \omega^2_s(\E)}.
\end{equation}
Note that Eq.~(\ref{eq:matrix_lebedev}), referred to below as the Lebedev equation, is the general equation since no approximations were used so far. The elements $G_{pk}$ depend on intensity parameter $\zeta$ as they are found after the stationary problem is solved.
The solution for $\Omega$ exists if the determinant of the following matrix is zero
\begin{equation}
    \label{eq:determinant}
    D(\Omega,\zeta) = \det \left|\delta_{p k} + \frac{\zeta}{h} \, G_{p k} (\Omega)\frac{Z_k(\Omega)/k}{\text{Im}Z/k} \right| =0.
\end{equation}

%%%%%%%%%%%%%%%%%%%%%%%%%%%%%%%%%%%%%%%%%%%%%%%%%%%%%%%%%%%
\subsection{Oide-Yokoya method}
%%%%%%%%%%%%%%%%%%%%%%%%%%%%%%%%%%%%%%%%%%%%%%%%%%%%%%%%%%%
In this section,  we re-derive, for completeness, another matrix equation that allows finding the bunch modes. According to this method  \cite{Oide_Yokoya1990}, called later the Oide-Yokoya,
the perturbed distribution function $\TF$ is expanded as
\begin{equation}
    \label{eq:oide_yokoya}
    \TF(\E,\psi,t) = \TF(\E,\psi,\Omega)\, e^{-i\Omega t} = e^{-i\Omega t} \sum_{m = 1}^{\infty} \left[C_m(\E,\Omega) \cos m\psi + S_m (\E, \Omega) \sin m\psi  \right].
\end{equation}
Inserting this expansion into Eq.~(\ref{eq:vlasov_equation_new})~\cite{Oide1994}, multiplying both sides by $\cos m\psi$ or $\sin m\psi$ and integrating over $\psi$,  the following system of equations can
be obtained
\begin{align}
    -i\Omega C_m(\E,\Omega) + m\omega_s(\E) S_m(\E,\Omega) &=
    \frac{m \omega_s(\E)}{\pi} \frac{d\F}{d\E}  \int_{-\pi}^{\pi}d\psi \, \Tilde{U}_\mathrm{ind}(\E,\psi,\Omega) \sin m\psi \nonumber \\
    &= 0, \label{eq:OY_eq1} \\
    -i\Omega S_m(\E,\Omega) - m\omega_s(\E) C_m(\E,\Omega) &= -\frac{m \omega_s(\E)}{\pi} \frac{d\F}{d\E}  \int_{-\pi}^{\pi}d\psi \, \Tilde{U}_\mathrm{ind}(\E,\psi,\Omega) \cos m\psi \nonumber\\
    &= 2 \zeta m \omega_s(\E)  \, \frac{d\F}{d\E}\sum_{k=-\infty}^{\infty}\frac{Z_k(\Omega)/k}{\text{Im}Z/k} \, \Tilde{\lambda}_k(\Omega) I_{mk}(\E) \label{eq:OY_eq2},
\end{align}
where the symmetry of $\Tilde{U}_\mathrm{ind}$ visible in Eq.~(\ref{eq:ind_potential_perturb})
was also used. 
Inserting expansion (\ref{eq:oide_yokoya}) into Eq.~(\ref{eq:line_density_pert_harm}), the perturbed line density harmonics can be presented in the form
\begin{equation}
    \label{eq:line_density_pert_harm_OY}
    \Tilde{\lambda}_k (\Omega)= \frac{\omega^2_{s0}}{h} \sum_{m =1 }^{\infty} \int_0^{\E_{\max}}d\E \frac{C_m(\E,\Omega) I^*_{mk}(\E)}{\omega_s(\E)}. 
\end{equation}
Combining  Eqs.~(\ref{eq:OY_eq1}-\ref{eq:line_density_pert_harm_OY}), one finally gets 
\begin{align}
    \label{eq:integral_eq_OY}
    \left[\Omega^2 - m^2\omega^2_s(\E) \right]C_m(\E,\Omega) &= 2 i \zeta \omega^2_{s0}\;   m^2 \omega^2_s(\E) \frac{d\F(\E)}{d\E} \nonumber\\
    &\times \sum_{m^\prime =1 }^{\infty} \int_0^{\E_{\max}}\frac{d\E^\prime}{\omega_s(\E^\prime)} \sum_{k=-\infty}^{\infty}\frac{Z_k(\Omega)/k}{h\text{Im}Z/k} I_{mk}(\E) I^*_{m^\prime k}(\E^\prime) C_{m^\prime}(\E^\prime,\Omega).
\end{align}
In the original work~\cite{Oide_Yokoya1990}, the functions $C_m(\E,\Omega)$ were defined as a combination of the step-like functions $s_n$, $C_{m}(\E,\Omega) = \sum_{n = 0}^{N_{\E}} s_n(\E) C_m(\E_n,\Omega) $, with
\[
s_n (\E)= \left\{
  \begin{array}{ll}
     1/\Delta \E_n, &  \E_n - \Delta \E_n/2 < \E \leq \E_n + \Delta \E_n/2 \\
    0, & \text{elsewhere,}
  \end{array}\right.
\]
where $\E_n$ is the $n$-th mesh point on the energy grid (in our calculations, it is assumed to be uniform  with the total number of points $N_\E$, i.e. $\Delta \E_n = \E_{\max}/N_{\E}$).
Thus, we obtain an eigenvalue problem of a linear algebra:
\begin{equation}
    \label{eq:eigen-system}
    \Omega^2 C_m(\E_n,\Omega) = \sum_{n^{\prime} = 1}^{N_{\E}} \sum_{m^\prime = 1}^{m_{\max}} M_{n m n^\prime m^\prime} \, C_{m^\prime}(\E_{n^\prime},\Omega),
\end{equation}
with the matrix elements
\begin{align}
    M_{n m n^\prime m^\prime} &= m^2 \omega^2_s(\E_n) \delta_{nn^\prime} \delta_{m m^\prime}  \nonumber\\ &-  \frac{2 \zeta m^2 \omega^2_s (\E_n)\omega_{s0}  \E_{\max}}{\pi A_N\omega_s(\E_{n^\prime}) N_{\E}} \frac{d g}{d\E}(\E_n)\,\text{Im} \left\{ \sum_{k=1}^{k_{\max}}\frac{Z_k/k}{h\text{Im}Z/k} I_{mk}(\E_n) I^*_{m^\prime k}(\E_{n^\prime}) \right \},\label{eq:matrix_OY}
\end{align}
$\delta_{n n^\prime}$ is the Kronecker delta, $m_{\max}$ is the maximum value of the azimuthal mode number, and $k_{\max}$ is the maximum value of the revolution harmonic number. It was shown in Ref.~\cite{Dyachkov} that approximating the integration over energy in Eq.~(\ref{eq:integral_eq_OY}) by a sum leads to the same eigenvalue problem. In general, all azimuthal modes and all frequency harmonics need to be included in calculations, i.e. $m_{\max}\to \infty$ and $k_{\max} \to \infty$. In practice, a truncation is used that might affect the results, which will be discussed in the following section. Note, that in the matrix element $M_{n m n^\prime m^\prime}$ dependence of impedance $Z_k$ on  $\Omega$ was neglected, so that an eigenvalue problem can be solved.
%%%%%%%%%%%%%%%%%%%%%%%%%%%%%%%%%%%%%%%%%%%%%%%%%%%%%%%%%%%
\section{\label{sec:Thesholds} Loss of Landau Damping}
%%%%%%%%%%%%%%%%%%%%%%%%%%%%%%%%%%%%%%%%%%%%%%%%%%%%%%%%%%%

In this section, we will consider the LLD thresholds in a general case and in application to the LHC, whose main parameters are listed in Table~\ref{tab:LHC_parameters}. 
For the LHC examples, with the beam being above transition energy, we will focus on a non-accelerating case with $\phi_{s0} = \pi$, and use the rf voltage that corresponds to the operational value at the flat bottom. 
\begingroup
\begin{ruledtabular}
\begin{table}[b!]
	\caption{The machine and rf parameters of the LHC~\cite{LHCDR}. 
	}
	\begin{center}
		\begin{tabular}{l  c  c  c  c }
			Parameter & Units & Flat bottom & Flat top \\
			\hline
			Circumference, $C$ & m & \multicolumn{2}{c}{26658.86}\\
			Harmonic number, $h$&   & \multicolumn{2}{c}{35640} \\
			Transition Lorentz factor, $\gamma_\mathrm{tr}$ &  & \multicolumn{2}{c}{55.76} \\
			rf frequency, $f_\mathrm{rf}$& MHz &\multicolumn{2}{c}{400.79} \\
			Beam energy, $E_0$ &TeV & 0.45 & 6.5\\
			rf voltage amplitude, $V_0$ & MV  & 6 & 12\\
			Estimated inductive impedance, Im$Z/k$ & $\Omega$ & 0.07& 0.076\\
		\end{tabular}
	\end{center}
	
	\label{tab:LHC_parameters}
\end{table}
\end{ruledtabular}
\endgroup

Analyzing solutions of the dispersion integral obtained from the Vlasov equation for infinite plasma,
N. G. van Kampen~\cite{vKampen1, vKampen2}  
has found that they have a continuous and
a discrete part.
In our case, at low intensities ($\zeta\approx 0$),
all van Kampen modes belong to the continuous spectrum, $\Omega=m\omega_{s}(\E)$, and the corresponding eigenfunctions are singular.
Landau damping results then from the phase mixing of these modes, which do not represent the collective motion of the particles. 
Above the threshold 
the discrete van Kampen modes can emerge from the continuous spectrum, implying that Landau damping is lost~\cite{YHChin1983}. 
These modes are described by regular functions and by definition they lie outside $\omega_{s}(\E)$.
The Oide-Yokoya method~\cite{Oide_Yokoya1990} used to calculate numerically the van Kampen modes allowed to determine the LLD threshold in different situations~\cite{burov2012van}. 

%%%%%%%%%%%%%%%%%%%%%%%%%%%%%%%%%%%%%%%%%%%%%%%%%%%%%%%%%%%%%%%%%%%%%
\subsection{\label{subsec:analytic_derivation} The LLD threshold for constant inductive impedance}
%%%%%%%%%%%%%%%%%%%%%%%%%%%%%%%%%%%%%%%%%%%%%%%%%%%%%%%%%%%%%%%%%%%%%
\subsubsection{Analytic criteria}
Here we first derive an analytic expression for the LLD threshold using the Lebedev equation~(\ref{eq:matrix_lebedev}). Then it will be compared with the results of the  semi-analytic calculations using both matrix equations described in Section~\ref{sec:perturbation} as well as with the macro-particle simulations.
In what follows we will focus on the case of $\eta \text{Im} Z/n >0$ (space charge below transition or inductance above transition), while the opposite sign is briefly discussed in Sec.~\ref{sec:discussions}. 

For the dipole mode ($m=1$), the LLD threshold is reached
when maximum 
eigenvalue $\Omega_{1n}$ of the matrix~(\ref{eq:matrix_OY}) equals the maximum incoherent frequency $ \hat{\omega}_s = \max[\omega_s(\E)]$, i.e. $\max(\Omega_{1n}) = \hat{\omega}_s$.
At low intensities, the synchrotron frequency distribution in a single rf system is a monotonic function of the energy of synchrotron oscillations $\E$. 
Assuming that still holds at the LLD threshold for a dipole mode $m=1$, we can search for the bunch intensity, at which $\Omega = \hat{\omega}_s = \omega_{s}(0)$ is a solution of Eq.~(\ref{eq:matrix_lebedev}).
Since, as follows from Eq.~(\ref{eq:Imk}), $I_{mk}(0) = 0$, the integral (\ref{eq:BTM_new}) defining the elements $G_{pk}$ converges for all $p$ and $k$.

To find the 
solution of Eq.~(\ref{eq:determinant}), we can use the following property of the matrix
\begin{equation}
    \label{eq:matrix_exp}
    \text{det}\left[\exp\left(\varepsilon\; X\right)\right] = \exp\left[\varepsilon\;\text{tr}\left(X\right)\right],
\end{equation}
where $\text{tr}(X)$ is the trace of an arbitrary square matrix $X$, and $\varepsilon$ is the small parameter $\varepsilon\ll 1$. Expansion of the exponent up to the first order of $\varepsilon$ yields,
\begin{equation}
    \label{eq:matrix_det}
    \text{det}(I+\varepsilon\; X) \approx 1+\varepsilon\;\text{tr}(X),
\end{equation}
with the identity matrix $I$.
Thus, we get a general expression for the LLD threshold
\begin{equation}
    \label{eq:LLD_threshold_general}
    \zeta_\mathrm{th} = -h\left[\sum_{k=-\infty}^{\infty} G_{k k} (\Omega)\frac{Z_k(\Omega)/k}{\text{Im}Z/k}\right]^{-1}.
\end{equation}
Naturally, the parameter $\varepsilon \propto \zeta$, while its dependence on the bunch length will be deduced below. Thus, potential well distortion has to be neglected in calculation of $G_{kk}$ as it would be a higher-order term of the parameter $\varepsilon$.

The element $G_{kk}$ can be calculated analytically if
we consider short bunches ($\E_{\max}\ll 1$) in a single rf system. Then, $\phi(\E,\psi)\approx \sqrt{2\E} \cos\psi$, $\omega_s(\E) \approx \omega_{s0}(1 - \E/8)$, and the functions $I_{mk}(\E)$ can be approximated by Bessel functions $J_m(x)$ of the first kind and the order $m$,
\begin{equation}
\label{eq:Imk_approx}
I_{m k} (\E) \approx i^m J_m\left(\frac{k}{h}\sqrt{2\E} \right).
\end{equation}
If only the first azimuthal mode ($m=1$) is taking into account, from Eq.~(\ref{eq:BTM_new}) we can obtain $G_{kk}$ at the LLD threshold ($\Omega = \omega_{s0}$) 
 
\begin{equation}
    \label{eq:Gkk}
    G_{kk} = -\frac{8i}{\pi A_N \phi^2_{\max}}\int^1_0 dx \frac{dg(x)}{dx} J^2_1\left(\frac{kx}{h}\phi_{\max} \right)\frac{ 1-\phi^2_{\max}x^2/16}{x^2 - \phi^2_{\max} x^4/32}.
\end{equation}
Here we used a new variable $x = \sqrt{\E/\E_{\max}}$ and defined the maximum phase amplitude $\phi_{\max} = \sqrt{2\E_{\max}}$. 
Below, we will consider the distributions of the binomial family $g(x) = \left(1-x^2\right)^\mu$ for which the normalization factor $A_N = \phi^2_{\max}/(2\mu+2)$ can be found using Eq.~(\ref{eq:normalization_factor}), once the synchrotron frequency spread is neglected. Then $G_{kk}$ is
\begin{align}
    G_{kk} &\approx i\frac{32\mu(\mu+1)}{\pi \phi^4_{\max}}\int^1_0 dx \frac{\left(1-x^2\right)^{\mu-1}}{x} J^2_1\left(\frac{kx}{h}\phi_{\max} \right) \nonumber \\
    &= i\frac{16  \mu(\mu+1)}{\pi \phi^4_{\max} }\left[1-  {}_1 F_2 \left(\frac{1}{2}; 2, \mu ; -{}y^2 \right)\right],\label{eq:Gkk1}
\end{align}
where ${}_p F_q(a_1,..,a_p; b_1,...,b_q;z)$ is the generalized Hypergeometric function and ${}y = k \phi_{\max}/h$. These matrix elements can be presented as a combination of Bessel functions for the particular values of $\mu$. For example, in the case of $\mu = 1/2$ and $\mu=2$ discussed below, one obtains $G_{kk}\propto \left[1-J_1(2{}y)/{}y\right]$ and $G_{kk}\propto \left[1/2-J^2_0({}y) -J^2_1({}y)+J_0({}y)J_1({}y)/{}y\right]$,
respectively. 
As $G_{kk}\propto 1/\phi_{\max}^4$, we can define the small parameter
\[\varepsilon = \zeta/\phi^4_{\max}\]
and check later validity of expansion~(\ref{eq:matrix_det}).

\begin{figure}[tb!]
	\centering
	\includegraphics[width = 0.6\textwidth]{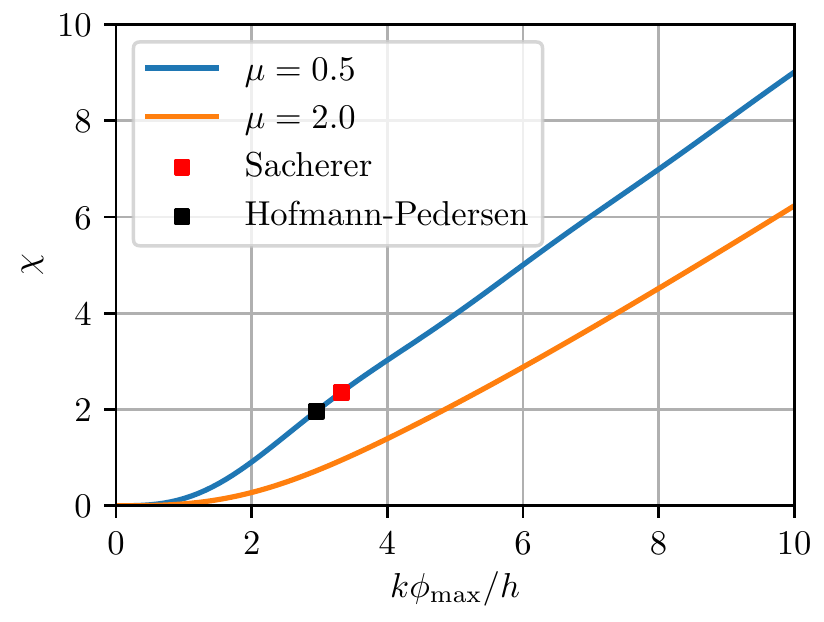}
	\caption{Examples of the function $\chi(k\phi_{\max}/h,\mu)$ defined by Eq.~(\ref{eq:chi_function}) for two different distribution functions from the binomial family ($\mu = 0.5, 2$). The values $k\phi_{\max}/h$ for    which the LLD thresholds correspond effectively to that used in Sacherer \cite{Sacherer2} and Hofmann-Pedersen \cite{Hofmann1979} criteria are shown with squares.}
	\label{fig:chi_function}
\end{figure}

Considering the constant inductive impedance $Z_k = i k \;\text{Im} Z/k$, the sum in Eq.~(\ref{eq:LLD_threshold_general}) can also be analytically evaluated by approximating it with an integral
\[
\frac{1}{h}\sum_{k=-\infty}^{\infty} G_{k k} (\Omega)\frac{Z_k(\Omega)/k}{\text{Im}Z/k} \approx \frac{i}{h}\int_{-\infty}^{\infty} dk G_{k k} (\Omega) \to \infty,
\]
which diverges, except the case of $\mu=0$ (a water-bag distribution).  
The derivative of this distribution function $\F^{\prime}\propto \delta(\E-\E_{\max})$. Thus the integrals in Eq.~(\ref{eq:BTM}) can be  evaluated analytically (without any approximations), yielding that each element $G_{pk}(\Omega)$ is proportional to $\omega_s(\E_{\max})$. Since, $\omega_s(\E_{\max}) = 0$ due to the potential well distortion (see Appendix~\ref{annex:B}), the determinant $D(\Omega,\zeta)$ is never zero and the LLD threshold does not exist.
For other values of $\mu$, once we have truncated the sum at some $k_{\max}$, the LLD threshold is
\begin{equation}
    \label{eq:threshold_binom_phimax}
    \zeta_\mathrm{th} = \frac{\pi \phi^5_{\max}}{32\mu(\mu+1)\chi({}y_{\max},\mu)} ,
\end{equation}
or in terms of intensity
\begin{equation}
    \label{eq:threshold_binom_phimax_int}
    N_{p,\mathrm{th}} = -\frac{\pi V_0\cos{\phi_{s0}} \phi^5_{\max}}{32 q h^2\omega_0  \mu(\mu+1)\chi({}y_{\max},\mu)\mathrm{Im}Z/k} ,
\end{equation}
where ${}y_{\max} = k_{\max} \phi_{\max}/h$ and we introduced the function
\begin{equation}
    \label{eq:chi_function}
    \chi({}y,\mu) ={}y\left[1 - {}_2 F_3 \left(\frac{1}{2},\frac{1}{2}; \frac{3}{2},2, \mu; -{}y^2 \right) \right].  
\end{equation}
Examples of this function for two different $\mu$ values are shown in Fig.~\ref{fig:chi_function}.

For $\mu=1/2$, 
the Sacherer formalism proposes  the LLD threshold (in our notations)~\cite{NG}
\[
\zeta_1 = \phi^5_{\max}/18
\]
Similarly, the Hofmann-Pedersen approach~\cite{Hofmann1979} applied for short bunches yields, 
\[
\zeta_2 = \phi^5_{\max}/15.
\]
The same expressions can be obtained from Eq.~(\ref{eq:threshold_binom_phimax}) for ${}y_{\max} \approx 3.32 $ and ${}y_{\max} \approx 2.95 $, respectively, (see also Fig.~\ref{fig:chi_function}). This implies that to have these  thresholds, the impedance each time has to be truncated at the frequency  $f_c = f_0 k_{\max} \approx1/\tau_\mathrm{full}$, inversely proportional to the full bunch length $\tau_\mathrm{full} = 2\phi_{\max}/\omega_\mathrm{rf}$. This is a low-frequency approximation while the constant inductive impedance $Z_k/k$ does not decay for $k_{\max}\xrightarrow{}\infty$.

Since the generalized Hypergeometric function ${}_2 F_3$ approaches zero for ${}y \to \infty$ a simple expression for the LLD threshold can be obtained
\begin{equation}
    \label{eq:threshold_binom_phimax_high_freq}
    \zeta_\mathrm{th} \approx \frac{\pi \phi^4_{\max}h}{32\mu(\mu+1)k_{\max}}.
\end{equation}
Thus, we see that the threshold is inversely proportional to the cutoff frequency and the fifth power in the dependence on the bunch length is replaced by the fourth. Our analytic formula ($\varepsilon \ll1$) is justified for impedance with $k_{\max} \gg h/8$ (for $\mu \geq 1/2$).

\subsubsection{Calculations using code MELODY}

\begin{figure}[b!]
	\centering
	\includegraphics[width = 0.48\textwidth]{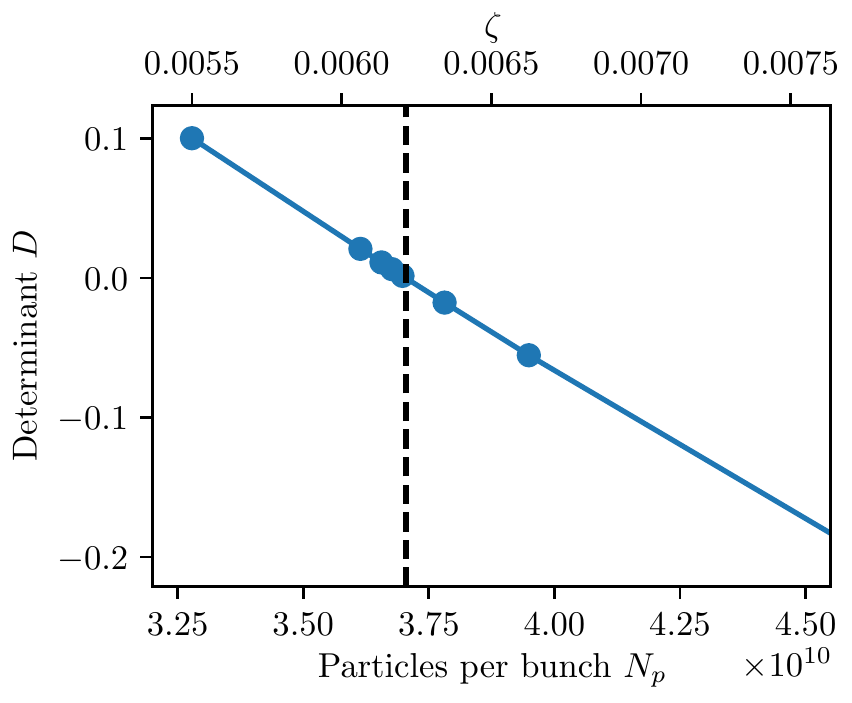}
	\includegraphics[width = 0.47\textwidth]{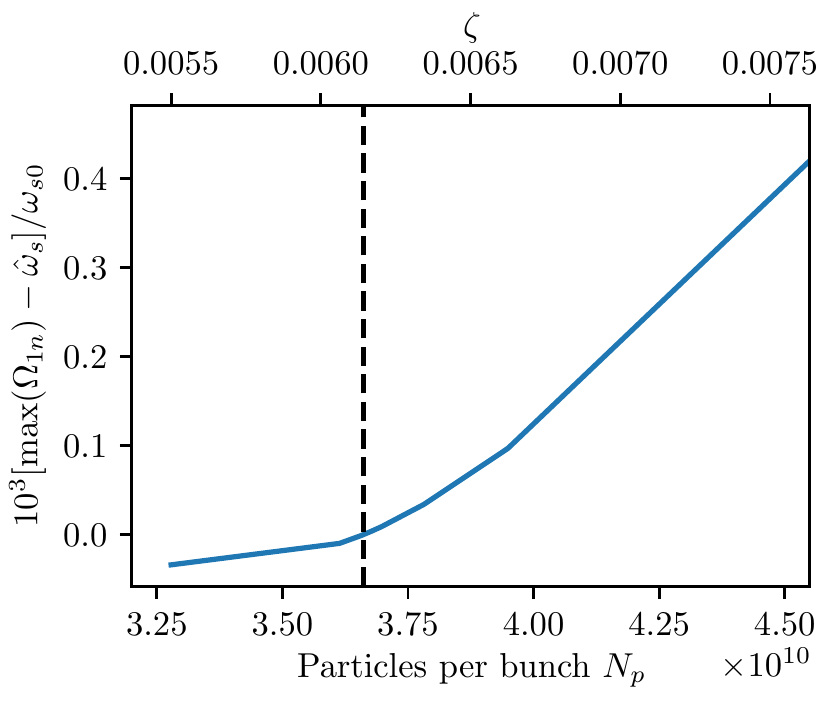}
	\caption{Examples of the LLD threshold evaluation (shown by a vertical dashed line) in MELODY using 
	the Lebedev equation (left) and 
	Oide-Yokoya method (right)   for $\eta > 0$, $\text{Im}Z/k= 0.07 \;\Omega$, $f_c = 4$~GHz, $V_0 = 6$~MV, $\mu = 2$, and zero-intensity  $\tau_{4\sigma}= 0.82$~ns.}
	\label{fig:threshold_identification}
\end{figure}

The analytic threshold~(\ref{eq:threshold_binom_phimax}) can now be compared with semi-analytical results obtained using code MELODY (Matrix Equations for LOngitudinal beam DYnamics calculations)~\cite{MELODY}.
In calculations based on the Lebedev equation~(\ref{eq:determinant}), the determinant $D(\hat{\Omega},\zeta)$ is numerically evaluated and the threshold $\zeta_\mathrm{th}$ corresponds to $D(\hat{\Omega},\zeta_\mathrm{th}) = 0$. This is illustrated in Fig.~\ref{fig:threshold_identification} (left) for the LHC parameters and $\mu=2$ observed in measurements~\cite{JMPhD}.

In Oide-Yokoya method, the eigenvalues of the matrix in Eq.~(\ref{eq:matrix_OY}) are calculated as a function of the intensity parameter $\zeta$.  To find the threshold, the difference between the maximum eigenfrequency and the maximum incoherent frequency is evaluated. The threshold corresponds to the zero of this function, as in an example in Fig.~\ref{fig:threshold_identification} (right).

\begin{figure}[t!]
% 	\centering
	\includegraphics[width = 0.6\textwidth]{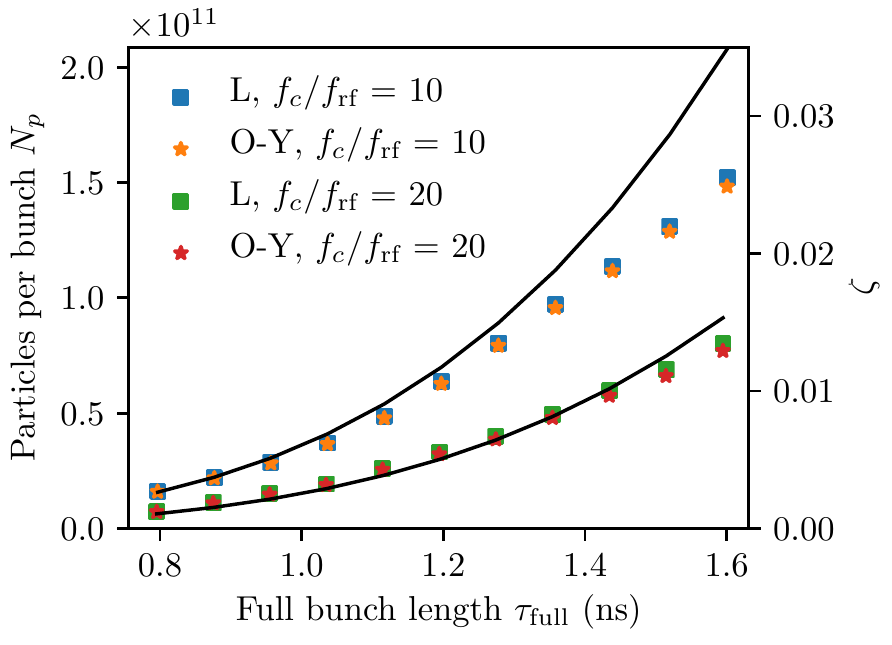}
	\caption{The LLD intensity threshold as a function of the full bunch length $\tau_\mathrm{full}$  calculated using the Lebedev equation (\ref{eq:matrix_lebedev}) and the Oide-Yokoya method (\ref{eq:matrix_OY}) for different cutoff frequencies ($f_c = k_{max}f_0$) of the inductive impedance. The analytic predictions from Eq.~(\ref{eq:threshold_binom_phimax}) are shown by solid lines. Case of $\eta > 0$ and other parameters are $\text{Im}Z/k= 0.07 \;\Omega$, $V_0 = 6$~MV, and  $\mu = 2$. The corresponding intensity parameter $\zeta$ is shown on the second vertical axis.
	}
	\label{fig:comp_threshold_ind}
\end{figure}

In Fig.~\ref{fig:comp_threshold_ind} we compare the LLD threshold as a function of the full bunch length  calculated for two different cutoff frequencies using analytic Eq.~(\ref{eq:threshold_binom_phimax}) and code MELODY. Indeed, the analytic expression derived from the Lebedev equation agrees well with the corresponding exact solution evaluated semi-analytically. As expected there is some discrepancy for larger bunch lengths since the analytic threshold was derived in short-bunch approximation while still taking into account the synchrotron frequency spread. Moreover, one can see that numerical results obtained from MELODY using the Oide-Yokoya method and the Lebedev equation agree almost perfectly, with the maximum 2\% relative error in the covered bunch-length range. We also observe that dependence on the bunch length is even slightly smaller than in the fourth power. 

\begin{figure}[t!]
% 	\centering
	\includegraphics[width = 0.6\textwidth]{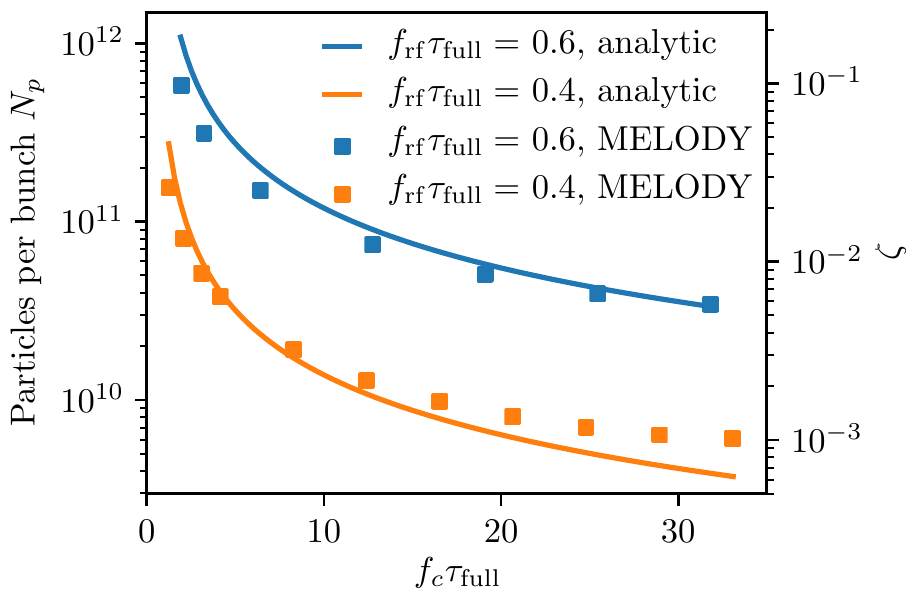}
	\caption{The LLD intensity threshold in the logarithmic scale as a function of the cutoff frequency of reactive impedance $f_c = k_{\max} f_0$ (multiplied by the full bunch length $\tau_\mathrm{full}$) for two different values of $\tau_\mathrm{full}$. The analytic predictions from Eq.~(\ref{eq:threshold_binom_phimax}) are shown by solid lines and the results of semi-analytic calculations using MELODY - by squares.
	Parameters as in Fig.~\ref{fig:comp_threshold_ind} The corresponding intensity parameter $\zeta$ is shown on the second vertical axis.
	}
	\label{fig:comp_threshold_ind_fcut}
\end{figure}

The semi-analytic results confirm that for a given bunch length, the threshold reduces as a function of the cutoff frequency (Fig.~\ref{fig:comp_threshold_ind_fcut}). The overall agreement between the analytical criterion~(\ref{eq:threshold_binom_phimax}) and the semi-analytic results is very good.

\begin{figure}[tb!]
	\centering
	\includegraphics[width = 0.6\textwidth]{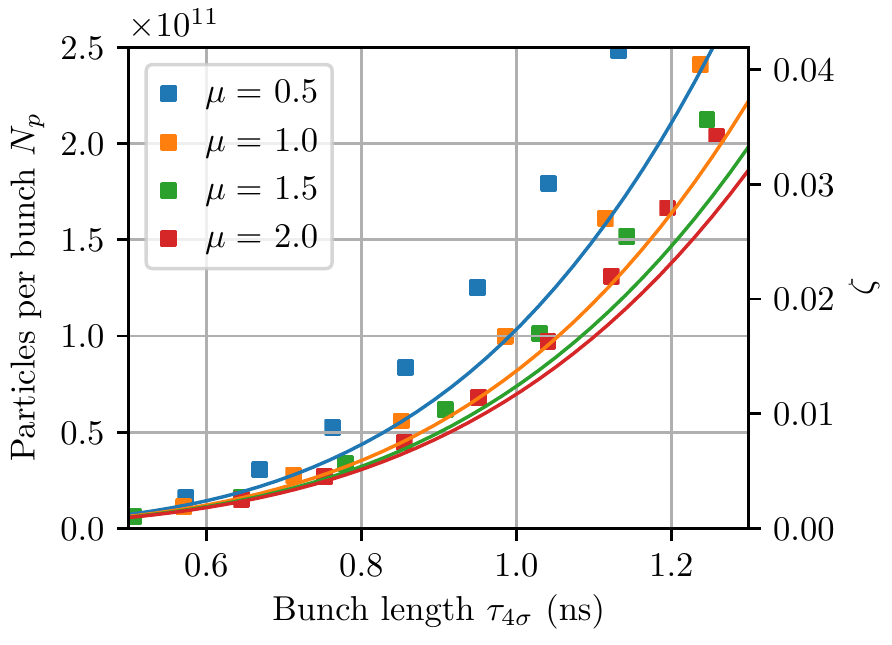}
	\caption{The LLD intensity threshold as a function of the bunch length for the binomial particle distribution with different values of $\mu$ calculated using  MELODY (squares) and Eq.~(\ref{eq:threshold_binom_phimax}) (lines) for $\eta>0$, $V_0 = 6$ MV and inductive impedance Im$Z/k= 0.07 \;\Omega$ with  $f_c = 4$~GHz.  The corresponding intensity parameter $\zeta$ is shown on the second vertical axis.}
	\label{fig:comp_threshold_ind_diff_mu}
\end{figure}

We also evaluated the LLD thresholds as a function of the scaled bunch length $\tau_{4\sigma}$ for a fixed cutoff frequency and different powers $\mu$ of the binomial distribution.  The analytic prediction~(\ref{eq:threshold_binom_phimax}) slightly underestimates the LLD threshold evaluated semi-analytically, 
see Fig.~\ref{fig:comp_threshold_ind_diff_mu}.
For $\mu>1$ the maximum relative error is below 30\% in the covered bunch-length range.
We see that threshold curves are getting closer for larger values of $\mu$, which is an important result showing a weak dependence on the details of distribution ones the scaled (FWHM) bunch length is used. 

%%%%%%%%%%%%%%%%%%%%%%%%%%%%%%%%%%%%%%%%%%%%%%%%%%%%%%%%%%%
\subsection{\label{subsec:BroadBandResonator} Effective impedance for estimation of the LLD threshold}
%%%%%%%%%%%%%%%%%%%%%%%%%%%%%%%%%%%%%%%%%%%%%%%%%%%%%%%%%%%
Very similar calculations can be done for any other impedance model. Indeed, the LLD threshold  obtained for the truncated constant inductive impedance [Eq.~(\ref{eq:threshold_binom_phimax})] can be applied for an arbitrary impedance $Z_k/k$ if we define the following effective impedance 
\begin{equation}
    \label{eq:effective_impedance}
    \left(\text{Im}Z/k\right)_\mathrm{eff} = \sum_{k = -k_\mathrm{eff}}^{k_\mathrm{eff}} G_{k k} \mathrm{Im}\left(Z_k/k\right) \bigg/ \sum_{k = -k_\mathrm{eff}}^{k_\mathrm{eff}} G_{k k}.
\end{equation}
where $k_\mathrm{eff}$ is the effective cutoff frequency, above which Im$Z_k/k$ is always negative, i.e. Im$Z_k/k<0$ for $k>k_\mathrm{eff}$.
Then, the LLD threshold becomes
\begin{equation}
    \label{eq:threshold_binom_phimax_eff}
    \zeta_\mathrm{th} = \frac{\pi \phi^5_{\max}}{32\mu(\mu+1)\chi(\mu,{}y_{\max})}\frac{\text{Im}Z/k}{ \left(\text{Im}Z/k\right)_\mathrm{eff}},
\end{equation}
and again in terms of intensity
\begin{equation}
    \label{eq:threshold_binom_phimax_eff_int}
    N_{p,\mathrm{th}} = -\frac{\pi V_0\cos{\phi_{s0}} \phi^5_{\max}}{32 q h^2\omega_0  \mu(\mu+1)\chi({}y_{\max},\mu)\left(\text{Im}Z/k\right)_\mathrm{eff}}.
\end{equation}
Here, in comparison to Eq.~(\ref{eq:threshold_binom_phimax_int}), $\mathrm{Im} Z/k$ is substituted by $\left(\text{Im}Z/k\right)_\mathrm{eff}$. Its applicability to an arbitrary impedance model is a subject of further studies, while in the present work we will focus on the  broad-band resonator impedance 
\begin{equation}
    \label{eq:bbr_impedance}
    Z^{\mathrm{bbr}}_k =\frac{ R }{1 + i Q \left(\frac{kf_0}{f_r} - \frac{f_r}{kf_0}\right)},
\end{equation}
where $R$ is the shunt impedance, $Q$ is the quality factor, and $f_r$ is the resonant frequency. To simplify comparison with the model of the constant inductive impedance, we choose $Q = 1$, $R= (\text{Im}Z/k)Qf_r /f_0$, with the same $\text{Im}Z/k = 0.07$~$\Omega$ as in the previous section while an impact of resonant frequency $f_r$ on the threshold is studied. 

\begin{figure}[tb!]
	\centering
	\includegraphics[width = 0.6\textwidth]{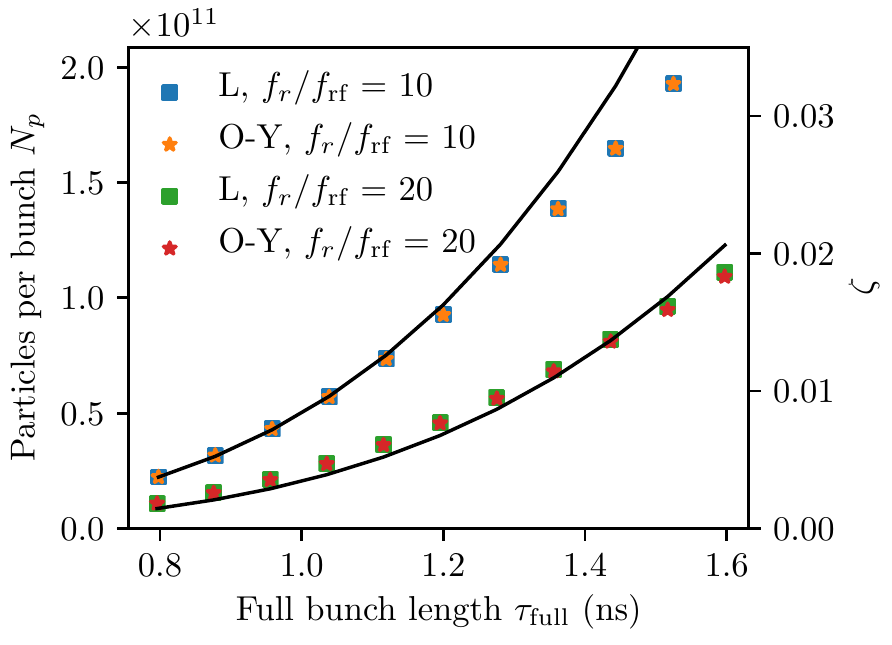}
	\caption{The LLD intensity threshold as a function of the full bunch length $\tau_\mathrm{full}$ calculated with MELODY using the Lebedev equation (squares) and the Oide-Yokoya method (stars) and analytical Eq.~(\ref{eq:threshold_binom_phimax_eff}) (lines) for a broad-band resonant impedance with $\text{Im}Z/k= 0.07 \;\Omega$, $Q=1$ and two different values of the resonant frequency $f_r$; $\eta>0$, $V_\mathrm{rf} = 6$ MV and  $\mu = 2$. The corresponding intensity parameter $\zeta$ is shown on the second vertical axis.}
	\label{fig:comp_threshold_ind_bbr}
\end{figure}

\begin{figure}[tb!]
	\centering
	\includegraphics[width = 0.6\textwidth]{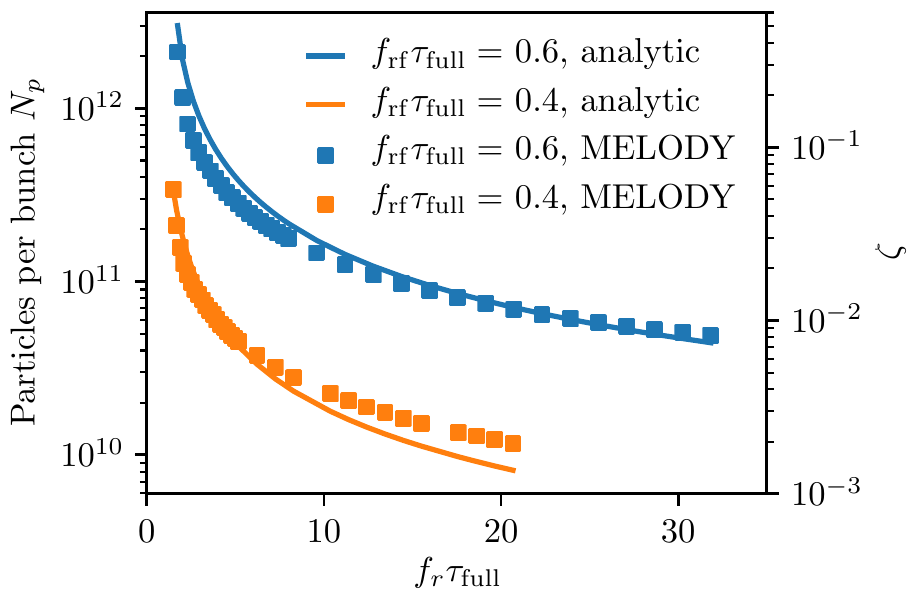}
	\caption{The LLD intensity threshold in the logarithmic scale as a function of $f_r \tau_\mathrm{full}$ calculated with MELODY (squares) and Eq.~(\ref{eq:threshold_binom_phimax}) (lines)  for  two different values of zero-intensity bunch length and a broad-band resonant impedance with $\text{Im}Z/k= 0.07 \;\Omega$ and $Q=1$; $\eta > 0$, $V_0 = 6$ MV  and $\mu = 2$. The corresponding intensity parameter $\zeta$ is shown on the second vertical axis.}
	\label{fig:comp_threshold_fr_tau}
\end{figure}

Figure~\ref{fig:comp_threshold_ind_bbr} shows dependence of the LLD threshold on the full bunch length for two values of the resonant frequencies $f_r$, similarly to Fig.~\ref{fig:comp_threshold_ind}. The analytical threshold given by Eq.~(\ref{eq:threshold_binom_phimax_eff}) agrees well with the results of semi-analytic calculations.
Dependence of the LLD threshold on the resonant frequency is also well reproduced for a given bunch length (Fig.~\ref{fig:comp_threshold_fr_tau}), as for the case of the truncated constant inductive impedance, see Fig.~\ref{fig:comp_threshold_ind_fcut}.
Note that these LLD thresholds were found taking into account only one azimuthal mode $m$. This is sufficient for  large $f_r \tau_\mathrm{full}$ values as the LLD intensity thresholds are lower for them and the contribution of higher-order azimuthal modes can be neglected. For smaller $f_r \tau_\mathrm{full}$, the contribution of higher-order azimuthal modes reduces the LLD threshold by a few percents.

The results of semi-analytic calculations were compared with 
the macro-particle simulations using the CERN Beam Longitudinal Dynamics code (BLonD)~\cite{BLOND2020} . We looked at the spectrum of the turn-by-turn data of different moments of the particle distribution depending on the azimuthal mode of interest (e.g., the mean value is sufficient to evaluate the frequencies of the dipole mode). 

\begin{figure}[tb!]
	\centering
	\includegraphics[width = 0.48\textwidth]{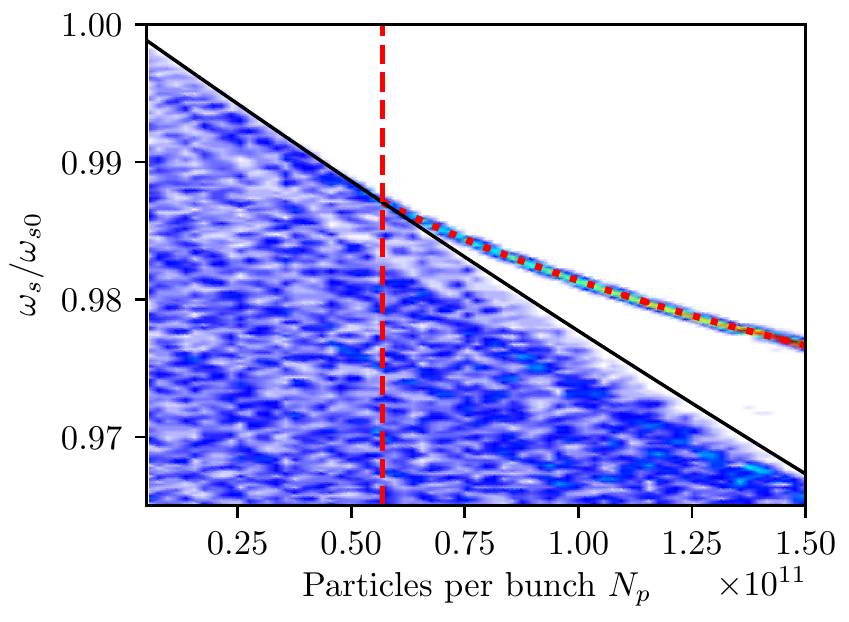}
	\includegraphics[width = 0.48\textwidth]{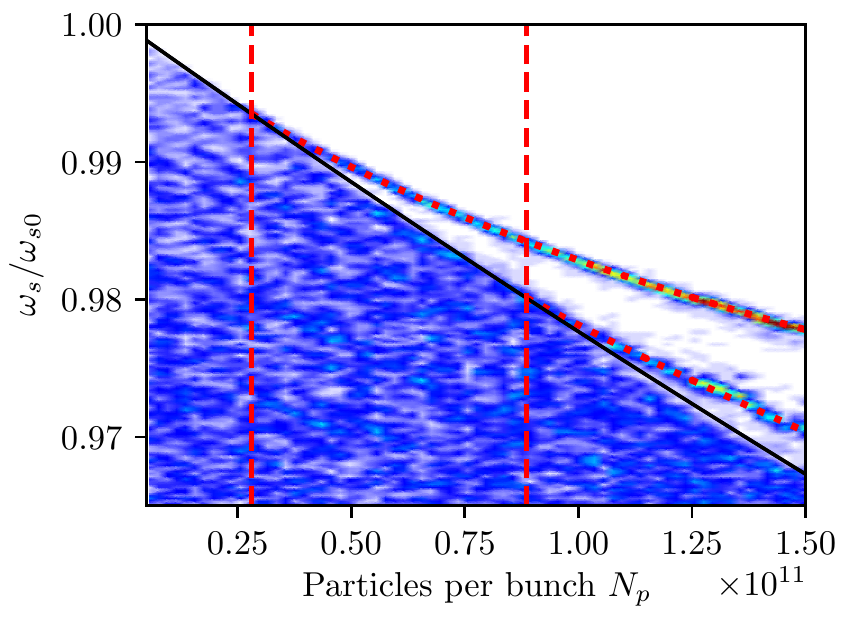}
	\caption{The real part of the normalized mode frequency found from macro-particle simulations using BLonD (blue color) and from MELODY (red dotted line) as a function of bunch intensity for a broad-band resonator impedance with Im$Z/k= 0.07 \;\Omega$, $Q=1$, $f_r = 4$~GHz (left) and $f_r = 8$~GHz (right). The maximum incoherent frequency obtained from MELODY is shown with black solid line. The dashed red lines indicate the LLD intensity thresholds. Calculations done for $\eta >0$, $V_\mathrm{rf} = 6$~MV, $\mu = 2$, zero-intensity $\tau_{4\sigma} =0.82$~ns, and $m \leq 5$.}
	\label{fig:comp_modes_B_vs_M_bbr}
\end{figure}

The bunch with $2\times 10^{6}$ macro-particles was generated, matched with intensity effects and then tracked for a sufficient number of turns to have a proper frequency resolution of the spectrum (typically up to $10^6$ turns). To evaluate induced voltage, we used at least 256 slices per rf bucket, that was sufficient to avoid uncontrolled emittance blowup due to numerical noise. Since at the LLD threshold, the mode frequency equals the maximum incoherent frequency it cannot be properly observed with a finite number of turns. However, above the threshold, the intensity dependence of the mode frequency can be accurately obtained and the van Kampen mode is seen as a strong peak above the incoherent band in  Fig.~\ref{fig:comp_modes_B_vs_M_bbr}.

Applying the Oide-Yokoya method in semi-analytic calculations by MELODY, the coherent modes above the threshold can be found in a standard way   as solutions of the eigenvalue problem. On the other hand, to find them using the Lebedev equation, we fix the intensity parameter $\zeta$ and vary $\Omega>\hat{\omega}_s$ until $D(\Omega,\zeta) = 0$. A good agreement between macro-particle simulations and semi-analytic calculations can be observed: as predicted the LLD threshold is lower for a higher resonant frequency. 
This is also the case for a higher-order radial mode of the dipole mode seen as the second emerged mode in Fig~\ref{fig:comp_modes_B_vs_M_bbr} (right). 

As it will be shown below, even though the LLD threshold is lower for impedance with higher cutoff or resonant frequencies, it does not necessarily mean that an initial perturbation will have a stronger impact on the beam at the threshold. 

%%%%%%%%%%%%%%%%%%%%%%%%%%%%%%%%%%%%%%%%%%%%%%%%%%%%%%%%%%%
\section{\label{sec:Impact_on_beam} Impact on the beam}
%%%%%%%%%%%%%%%%%%%%%%%%%%%%%%%%%%%%%%%%%%%%%%%%%%%%%%%%%%%
% \textcolor{red}{Subjects to include in this chapter}
% \begin{itemize}
%     \item Meaning of a rigid-dipole kick (injection error, phase modulation from RF)
%     \item Expansion of rigid-dipole mode using basis of eigenvectors
%     \item Mode shape in frequency and time domains below and above threshold; dependence on cutoff frequency
%     \item Bunch offset evolution after the kick
%     \item Depending on intensity fast decoherence, then depending on intensity either exponential damping or continuously oscillations (with or without beating)
%     \item Comparison with simulation (limitation due to a number of macro-particles)
%     \item Scaling of residual oscillation amplitude with bunch length ($\propto\tau^6$). It seems that if kick is normalized to the bunch length we get dependence $\propto \tau^5$
%     \item Comparison of one point from measurements in the LHC

% \end{itemize}
A rigid-bunch dipole perturbation (a kick) could be  a result of the injection phase or/and energy errors, but also of a phase noise in the rf system. 
%This is the situation when 
Evaluation of the beam response to a rigid-dipole perturbation is a common way to study the LLD in simulations and measurements. N.~G.~van~Kampen has shown that an arbitrary perturbation of the initial distribution behaves as a superposition of the waves that are solutions of the equation equivalent to Eq.~(\ref{eq:integral_eq_OY})~\cite{vKampen1}. 
% An expansion of an initial kick as a superposition of the bunch head-tail eigenmodes in the transverse plane~[Kornilov2012] was used to evaluate the dominating head-tail modes in the bunch offset evolution. 
Below we will use expansion of a rigid-dipole perturbation on the basis of van Kampen modes to obtain analytically its time evolution.
%the bunch-offset evolution after an initial perturbation. 
It was also used to describe the transverse oscillations of the colliding bunches~\cite{Alexahin1998}.
Note that a similar approach can be applied for the analysis of  higher-order perturbations (e.g., quadrupolar oscillations due to bucket mismatch).

\subsection{\label{subsec:RigidDipoleKick}Rigid-dipole kick as superposition of van Kampen modes}
As a result of a kick, the whole bunch has a phase  offset $\Delta \phi$ with the respect to its synchronous phase. Since $\lambda(\phi+\Delta\phi) = \lambda(\phi) + \Delta \phi \; d\lambda / d \phi$,  the rigid-bunch dipole mode can be described via the derivative of the line density $d\lambda / d \phi$.
Its spectral harmonics $\left(d\lambda/d\phi\right)_k$ can be obtained using Eq.~(\ref{eq:line_density_harm})
\begin{equation}
    \label{eq:rigid_harm}
    \left(\frac{d\lambda}{d\phi}\right)_k =\frac{ik}{h}\lambda_k = \frac{\omega_{s0}^2}{h}  \int_0^{\E_{\max}} d\E \, \frac{\F(\E)}{\omega_{s}(\E)} \frac{ik}{h} I_{0k}(\E).
\end{equation}
Dependence of harmonics on $kI_{0k}$ means that it is not a pure dipole ($m=1$) mode. The elements of the eigenvector of this mode can be approximated as $C_1^\mathrm{rd}(\E) \propto \sqrt{\E}\F^\prime (\E)$ and $C_m^\mathrm{rd}(\E) = 0$ for $m>1$~\cite{Oide1994}. We can define the harmonics of the approximate rigid-dipole mode, $\Tilde{\lambda}^{\mathrm{rd}}_k$, similar to Eq.~(\ref{eq:line_density_pert_harm_OY}),
\begin{equation}
    \label{eq:line_density_rda}
    \Tilde{\lambda}^{\mathrm{rd}}_k (\Omega)= \frac{\omega^2_{s0}}{h}  \int_0^{\E_{\max}}d\E \, \frac{C^\mathrm{rd}_1(\E) I^*_{1k}(\E)}{\omega_s(\E)}. 
\end{equation}
After finding eigenvalues and eigenvectors of matrix~(\ref{eq:matrix_OY}), this approximate mode can be expanded using the van Kampen modes as a basis
\begin{equation}
    \label{eq:mode_expansion}
    C_m^\mathrm{rd}(\E_n)= \sum_{m^\prime = 1}^{m_{\max}} \sum_{n^\prime = 1}^{N_{\E}} \alpha_{m^\prime n^\prime} C_{m}(\E_n,\Omega_{m^\prime n^\prime}),
\end{equation}
where $\alpha_{n^\prime m^\prime}$ is the expansion coefficient, $\Omega_{m^\prime n^{\prime}}$ is the van Kampen mode with the index $n^{\prime}$ that belongs to azimuthal mode $m^\prime$. We impose the following normalization of the eigenvectors 
\[
\sum_{m = 1}^{m_{\max}} \sum_{n = 1}^{N_\E} |C_m(\E_n,\Omega)|^2 = 1 \text{ and }  \sum_{m = 1}^{m_{\max}} \sum_{n = 1}^{N_\E}  |C_m^{\mathrm{rd}}(\E_n)|^2 = 1.
\]
Since the basis is not orthonormal, to find $\alpha_{nm}$, we construct a matrix from eigenvectors $K_{m n m^\prime n^\prime} = C_{m}(\E_n,\Omega_{m^\prime n^\prime})$, evaluate its inverse matrix $K^{-1}$, and finally obtain
\begin{equation}
    \label{eq:expansion_coeff}
    \alpha_{mn} = \sum_{m^\prime = 1}^{m_{\max}} \sum_{n^\prime = 1}^{N_\E} K^{-1}_{n m n^\prime m^\prime} C_{m^\prime}^\mathrm{rd}(\E_{n^\prime}).
\end{equation}

\subsection{\label{subsec:BunchOffsetAfterKick} Bunch offset evolution after a rigid-dipole kick}

Once the solution of the eigenvalue problem~(\ref{eq:matrix_OY}) and the mode expansion coefficients $a_{mn}$ from Eq.~(\ref{eq:expansion_coeff}) are found, the bunch offset evolution can be described as
\begin{equation}
    \label{eq:offset}
    \Delta \phi_b(t) =\kappa \sum_{m = 1}^{m_{\max}} \sum_{n = 1}^{N_\E} \alpha_{mn} \int_{-\phi_{\min}(\E_{\max})}^{\phi_{\max}(\E_{\max})} d\phi \; \phi \tilde{\lambda}(\phi,\Omega_{mn}) \cos(\Omega_{mn} t),
\end{equation}
where the factor $\kappa = \max \left[\lambda(\phi + \Delta \phi) - \lambda(\phi)\right]/\max\left[\lambda^\mathrm{rd}(\phi)\right]$ depends on the initial offset $\Delta \phi$. 
In what follows we will consider the broad-band resonator impedance model, while similar results can also be obtained for the truncated constant inductive impedance.

\begin{figure}[tb!]
	\centering
	\includegraphics[width = 0.6\textwidth]{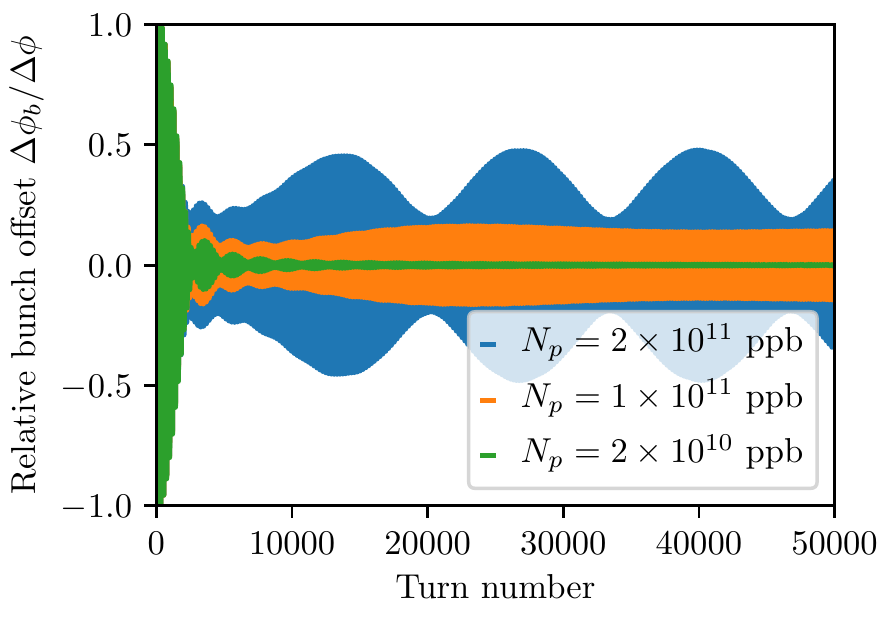}
	\caption{Offset evolution of bunch with different intensities  described by Eq.~(\ref{eq:offset}) after a kick of $\Delta \phi = 1$ deg for a broad-band resonant impedance with $\text{Im}Z/k= 0.07 \;\Omega$, $f_r = 4$ GHz and $Q =1$; $V_0 = 6$~MV, zero-intensity $\tau_{4\sigma} = 0.82$~ns, and $\mu = 2$. In this case, the calculated LLD threshold $N_\mathrm{th} \approx 5.7\times 10^{10}$.}
	\label{fig:kick_response}
\end{figure}

Considering a small initial kick $\Delta \phi \ll \omega_\mathrm{rf}\tau_{4\sigma}$ that does not cause a significant emittance blowup, we can evaluate analytically, using Eq.~(\ref{eq:offset}), the bunch evolution with time. The results for different intensities are shown in  Fig.~\ref{fig:kick_response}. 
The bunch oscillations are damped after the initial fast decoherence when intensity is lower than the LLD threshold. There are residual oscillations above the LLD intensity threshold and their amplitudes depend on intensity. At even higher intensities, one
typically observes a beating. 

\begin{figure}[tb!]
	\centering
	\includegraphics[width = 0.48\textwidth]{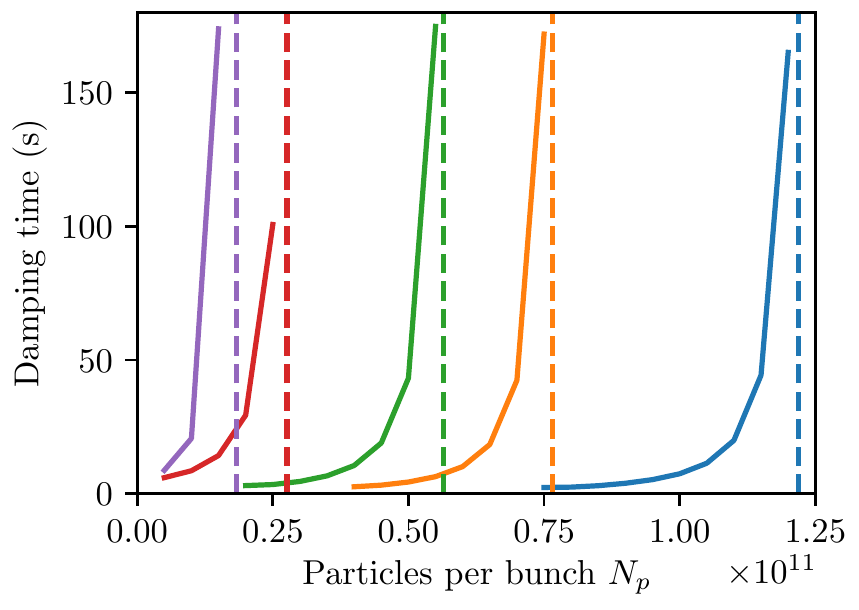}
	\includegraphics[width = 0.46\textwidth]{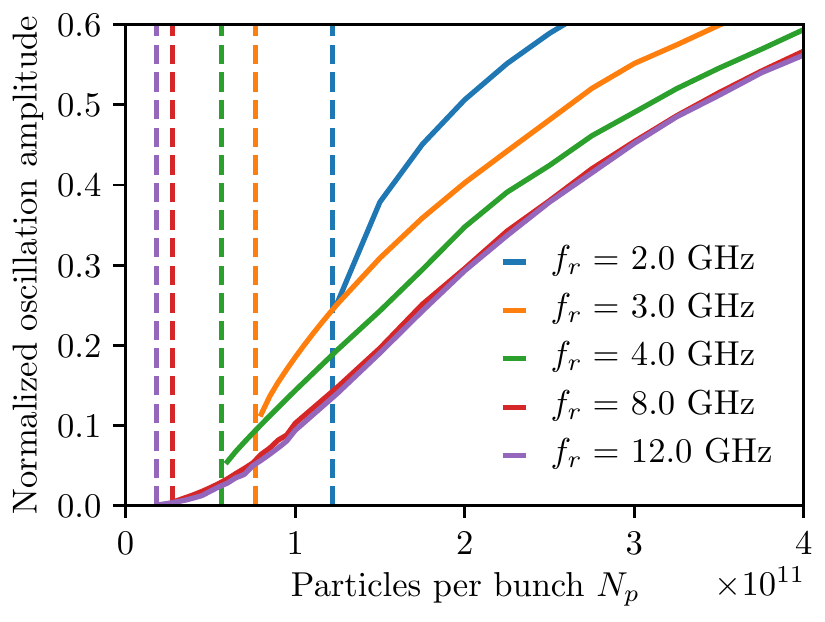}
	\caption{ The damping time below the LLD threshold (left) and the residual oscillation amplitudes normalised to the initial kick above the LLD threshold (right) versus bunch intensity for different resonator frequencies $f_r$. Other parameters are the same as in Fig.~\ref{fig:kick_response}. Dashed vertical lines indicate the corresponding LLD thresholds calculated with MELODY. 
	}
	\label{fig:kick_response_all}
\end{figure}

We can also see that the intensity dependency of the damping time is similar for different resonator frequencies (left plot in Fig.~\ref{fig:kick_response_all}). The damping time is short for $N_p\ll N_{p,\mathrm{th}}$, while it approaches infinity when the LLD threshold is reached. It means that Landau damping can be effectively lost even below the  threshold depending on the machine cycle length.

Moreover,  at the LLD threshold and above, the residual oscillation amplitude after a kick  depends on the resonant frequency. Comparing the average residual oscillation amplitudes, we see that,  at a given intensity, it is smaller for a higher $f_r$. We also observe that amplitude dependence on intensity saturates for $f_r \tau_{4\sigma} \gg 1$. These findings become more clear by analyzing the mode patterns in the frequency and time domains.

\begin{figure}[tb!]
	\centering
	\includegraphics[width = 0.48\textwidth]{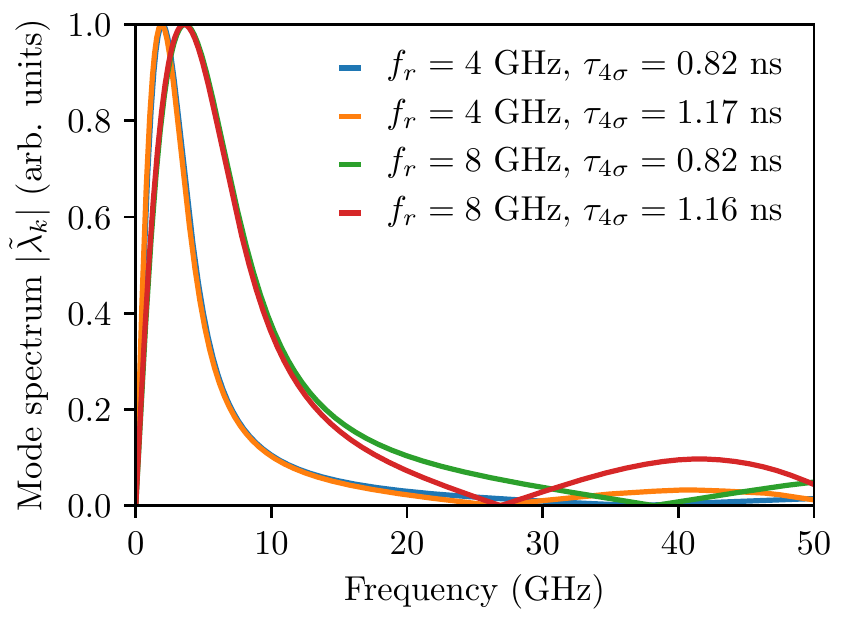}
	\includegraphics[width = 0.48\textwidth]{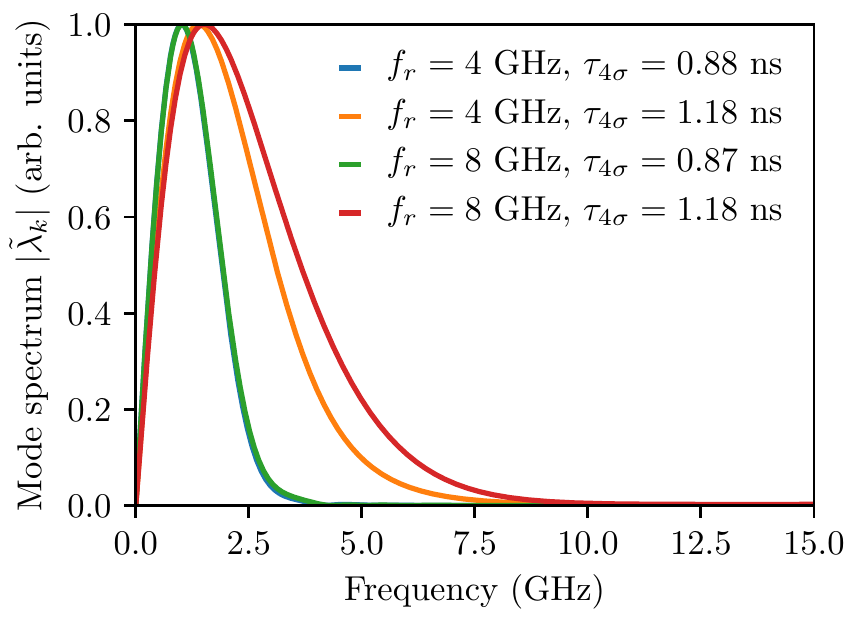}
	\caption{Coherent mode spectra at the LLD threshold (left) and 
	well above it (right) found using MELODY  for different values of resonator frequency $f_r$ and  bunch length $\tau_{4\sigma}$ for a broad-band resonator impedance with $\text{Im}Z/k= 0.07 \;\Omega$ and $Q=1$; $\eta>0$, $V_0 = 6$~MV   and $\mu =2$.}
	\label{fig:comp_mode_sp_diff_mu}
\end{figure}

Comparison of coherent mode spectra for different values of bunch length and resonant frequency in Fig.~\ref{fig:comp_mode_sp_diff_mu} (left)
shows that at the LLD threshold, the mode is not sensitive to details of particle distribution, but rather depends on impedance and its resonant frequency. This, in fact, explains why the LLD threshold significantly increases for $f_r \tau_\mathrm{full} \to 1$ as can be seen in  Fig.~\ref{fig:comp_threshold_fr_tau}. For sufficiently  low $f_r$, the characteristic width of the mode becomes larger than the bunch length, so the mode cannot exist inside the bunch.

\begin{figure}[tb!]
	\centering
	\includegraphics{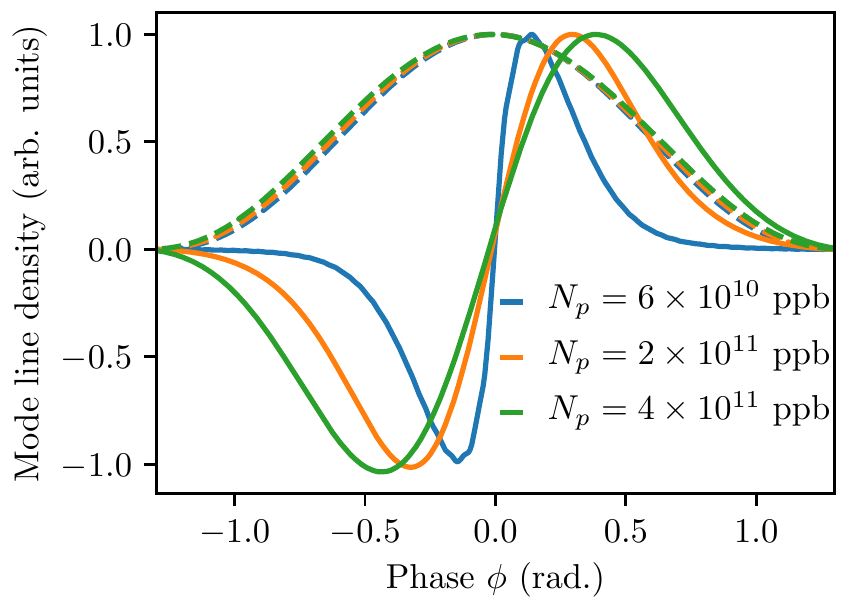}
	\caption{Coherent modes (solid lines) in time-domain calculated using MELODY for different bunch intensities and broad-band resonator model with $\text{Im}Z/k= 0.07 \;\Omega$, $Q=1$ and $f_r = 4$~GHz; $V_0 = 6$~MV and zero-intensity $\tau_{4\sigma}= 0.82$~ns. The corresponding stationary line densities (with $\mu = 2$) are shown by dashed lines. }
	\label{fig:comp_mode_td_diff_int}
\end{figure}

We also observe that the mode is ``wider" in the frequency domain for a higher  $f_r$, which would correspond to a more localized mode in the time domain. 
This explains the fact that beam response to the kick is stronger for lower $f_r$.
On contrary, if we compare the same modes well above the LLD threshold (at $N_p= 4\times10^{11}$ ppb, for example), they become very similar for the same bunch lengths but different $f_r$ (see right plot in Fig.~\ref{fig:comp_mode_sp_diff_mu}).
Typically, the coherent modes above the LLD threshold widen in the time domain as intensity increases (see examples in Fig.~\ref{fig:comp_mode_td_diff_int}), so the beam has a stronger response to a rigid-dipole offset as shown in Fig.~\ref{fig:kick_response_all}.

%%%%%%%%%%%%%%%%%%%%%%%%%%%%%%%%%%%%%%%%%%%%%%%%%%%%%%%%%%%%
\subsection{\label{subsec:Comp_sim_kick} Comparison with simulations and LHC measurements}
%%%%%%%%%%%%%%%%%%%%%%%%%%%%%%%%%%%%%%%%%%%%%%%%%%%%%%%%%%%%

\begin{figure}[tb!]
	\centering
	\includegraphics[width = 0.6\textwidth]{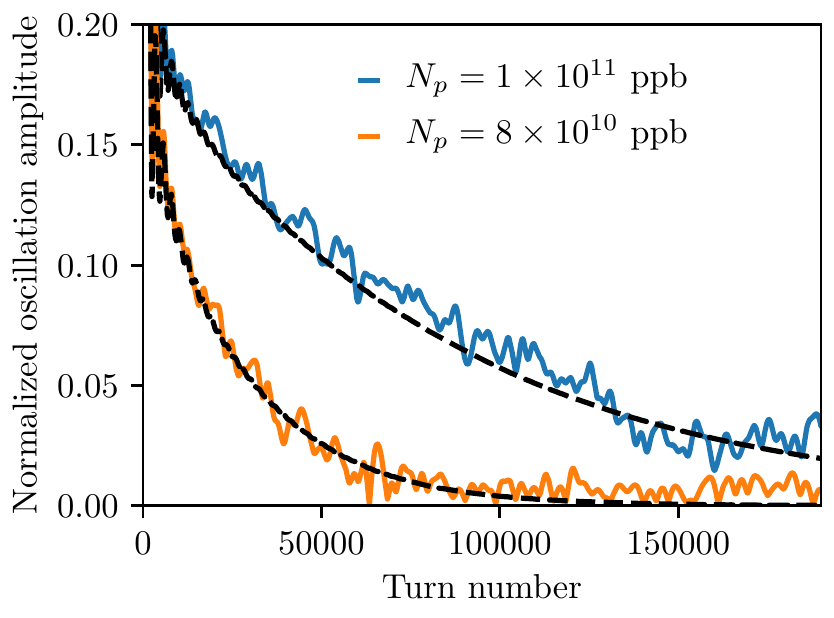}
 	\caption{Bunch offset  evolution  after a kick, found with MELODY using Eq.~(\ref{eq:offset}) (dashed lines) and from macro-particle simulations using BLonD (solid lines) for different bunch intensities below the LLD threshold $N_{p,\mathrm{th}} \approx 1.2 \times 10^{11}$ for a broad-band impedance with $\text{Im}Z/k= 0.07 \;\Omega$, $Q=1$ and $f_r = 2$~GHz; zero-intensity $\tau_{4\sigma} = 0.82$~ns, $\mu = 2$, and $V_0 = 6$~MV. }
	\label{fig:kick_response_blond_melody_damping}
\end{figure}

Below we will compare the results obtained so far from semi-analytic calculations using code MELODY with macro-particle simulations by code BLonD.
In simulations, the amplitude of oscillations from the turn-by-turn bunch offset was extracted using the Hilbert transform. \begin{figure}[tb!]
	\centering
	\includegraphics[width = 0.48\textwidth]{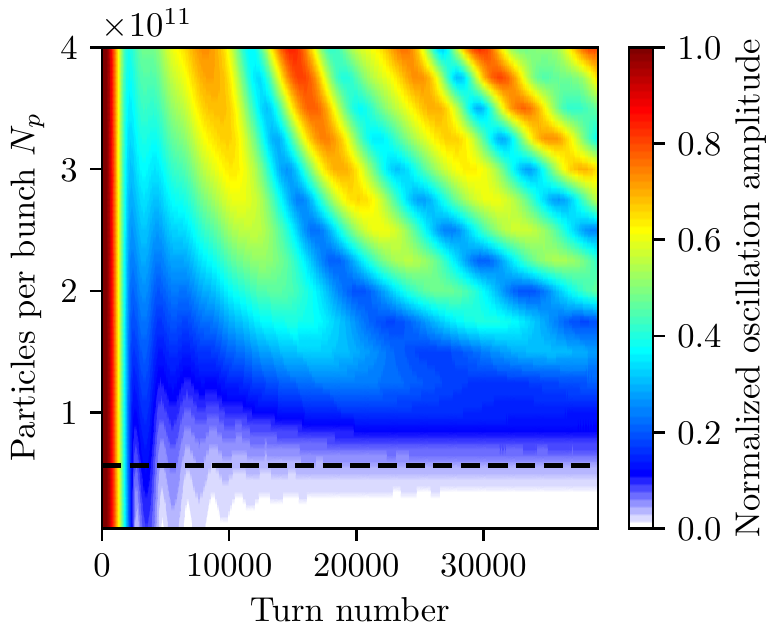}
	\includegraphics[width = 0.48\textwidth]{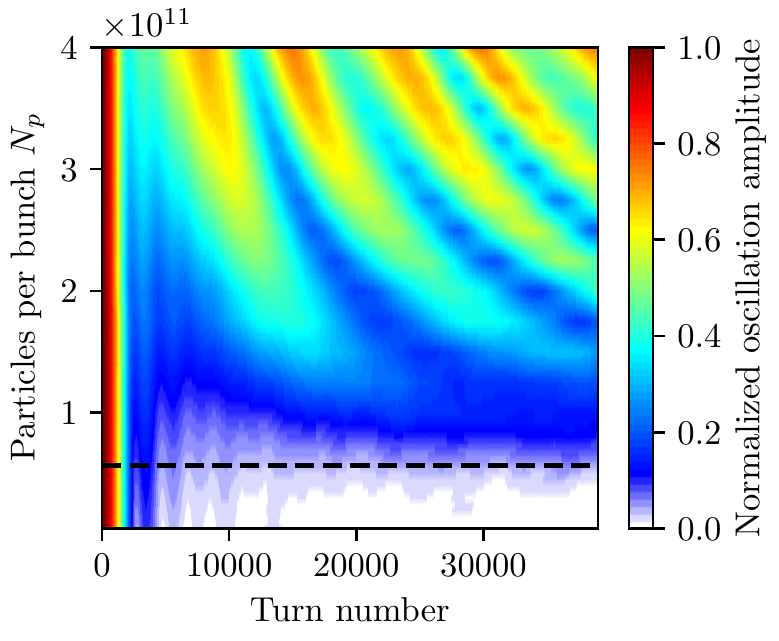}
	\caption{Time evolution of the bunch offset amplitude (color coding) after a kick, obtained from MELODY using Eq.~(\ref{eq:offset}) (left) and from macro-particle simulations using BLonD (right) for different intensities (vertical axes) and a broad-band impedance with  Im$Z/n= 0.07 \;\Omega$, $Q=1$, and $f_r = 4$~GHz; zero-intensity $\tau_{4\sigma} = 0.82$~ns, $\mu = 2$, and $V_0 = 6$~MV.
	The LLD threshold at $\approx 5.7\times 10^{10}$ is shown with a dashed line.}
	\label{fig:kick_response_blond_melody}
\end{figure}
An example of comparison  is shown in Fig.~\ref{fig:kick_response_blond_melody} for $f_r = 4$~GHz. 
One can see a good agreement above the  threshold, even though we used the approximate rigid-dipole mode in the analytic mode expansion. The exact phase of beating slightly differs, but the amplitudes are very similar. 
Below the threshold, where we expect a slow damping around the threshold,  in the simulations we observe the effect of noise even when $20\times 10^6$ macro-particles are used. 
To verify that damping is also present in the simulations,
we can use a lower resonant frequency, for example $f_r = 2$ GHz, since in this case the amplitude of the residual oscillations is larger than for higher resonant frequencies. We also increased the number of macro-particles up to $30\times 10^6$.
Indeed, as one can see in Fig.~\ref{fig:kick_response_blond_melody_damping}, BLonD simulations and MELODY calculations match very well even below the LLD threshold, though there is still some beating caused by the noise.

\begin{figure}[tb!]
	\centering
	\includegraphics[width = 0.48\textwidth]{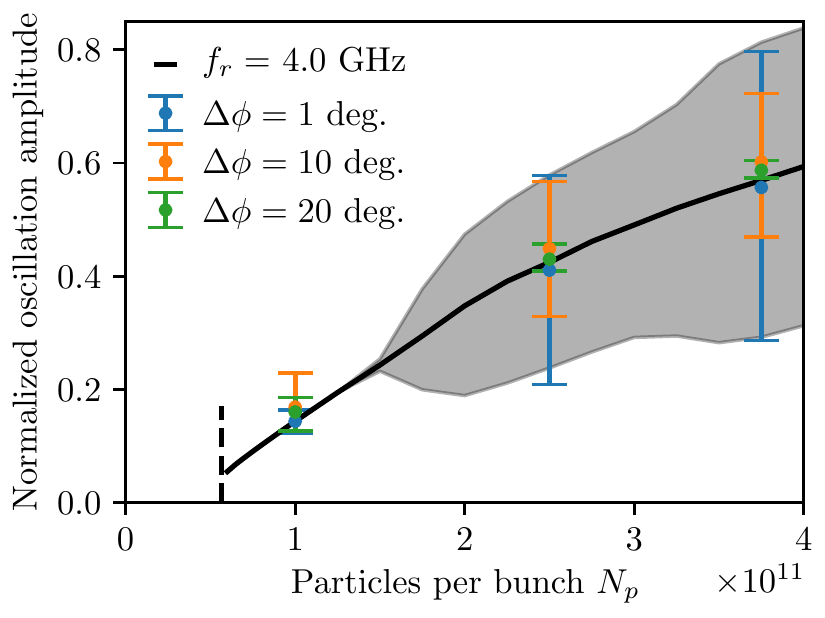}
	\includegraphics[width = 0.48\textwidth]{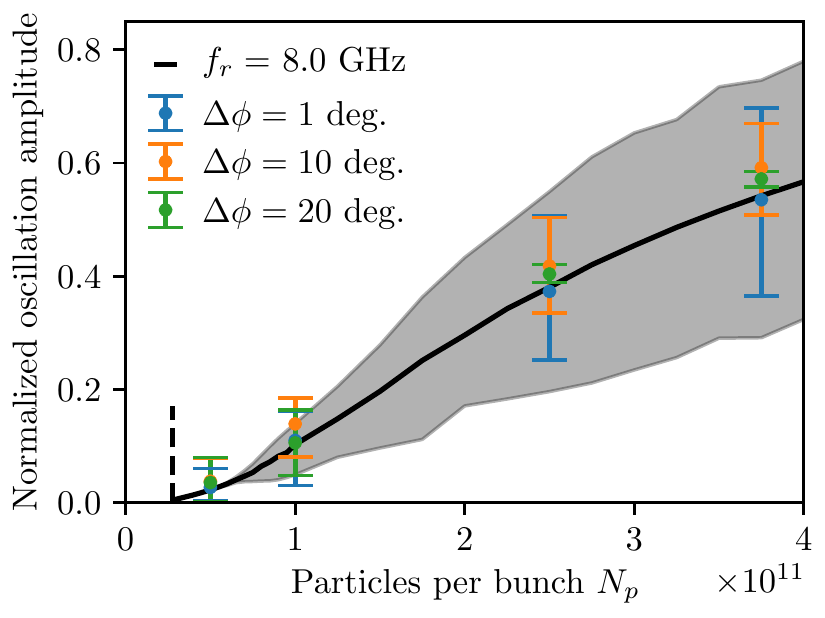}
	\caption{ Normalized average amplitude of residual oscillations after a kick of  different amplitudes obtained from BLonD macro-particle simulations (color points) and from MELODY using Eq.~(\ref{eq:offset}) (black curves) for different bunch intensities above the LLD threshold (vertical dashed lines). The shaded grey area and error-bars represent the minimum and maximum amplitudes, obtained from MELODY and BLonD, respectively. Results are for a broad-band impedance with $\text{Im}Z/k= 0.07 \;\Omega$, $Q=1$, $f_r = 4$~GHz (left) and $f_r = 8$~GHz (right); zero-intensity $\tau_{4\sigma} = 0.82$ ns, $\mu = 2$, and $V_0 = 6$~MV.}
	\label{fig:kick_strength}
\end{figure}

The BLonD simulations for different kick amplitudes are compared with results from MELODY in Fig.~\ref{fig:kick_strength}.
There is a good agreement for the minimum, mean, and maximum amplitude of oscillations for a small kick amplitude. The average oscillation amplitude can still be predicted even for larger kicks (up to 20 deg).

\begin{figure}[tb!]
	\centering
    \includegraphics[width = 0.6\textwidth]{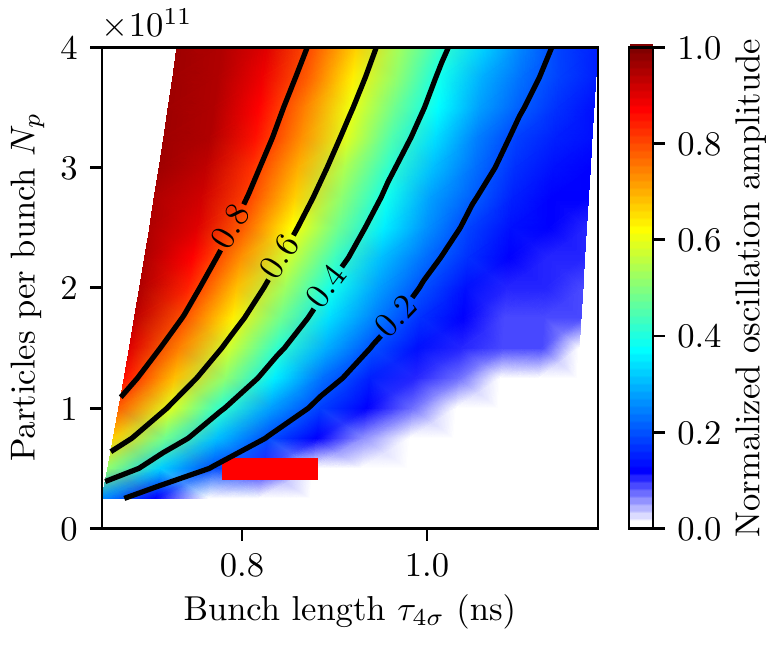}
	\caption{Residual oscillation amplitude (colors and black contour lines) after a kick of $\Delta \phi = 1$ deg. above the LLD threshold as a function of bunch intensity and the scaled bunch length for a broad-band impedance with $\text{Im}Z/k= 0.076 \;\Omega$, $Q=1$, and $f_r= 5$~GHz (the estimated cutoff frequency of the LHC beam pipe); $\mu =2$, $V_0 = 10$~MV. The parameters of LHC bunches with LLD  observed in measurements~\cite{JMPhD} are shown by a red box.}
	\label{fig:kick_response_map}
\end{figure}

Figure~\ref{fig:kick_response_map} shows a summary plot of the normalized oscillation amplitude after the one-degree kick as a function of the scaled bunch length $\tau_{4\sigma}$ and bunch intensity $N_p$. 
The chosen resonant frequency of 5 GHz corresponds to the cutoff frequency of the LHC vacuum chamber~\cite{LHCDR}. The rf voltage is    10~MV, as during the measurements in the LHC. We also denote the bunch parameters for those the LLD was observed~\cite{JMPhD}. One can obtain from the fit that the residual oscillation amplitude scales as $\tau_{4\sigma}^6$, while the LLD threshold is proportional to $\tau_{4\sigma}^4$.
The parameters from the measurements are in the region where Landau damping is lost, but residual amplitude is below 20\% of the kick amplitude, in agreement with observations. In the future, however, the LHC impedance model at high frequencies has to be revised to have more accurate predictions of the LLD threshold for the HL-LHC beam intensity.

%%%%%%%%%%%%%%%%%%%%%%%%%%%%%%%%%%%%%%%%%%%%%%%%%%%%%%%%%%%%
\section{\label{sec:discussions}Other cases: discussion}
%%%%%%%%%%%%%%%%%%%%%%%%%%%%%%%%%%%%%%%%%%%%%%%%%%%%%%%%%%%%
The present work was dealing so far with the binomial particle distributions and the inductive impedance above transition energy ($\eta \text{Im}Z/k>0$), which is typical for high energy colliders.
We have seen that the LLD threshold is decreasing with cutoff or resonant frequency of impedance. Here we will consider two cases,
when the LLD threshold does not depend on these frequencies.

%%%%%%%%%%%%%%%%%%%%%%%%%%%%%%%%%%%
%%%%%%%%%%%%%%%%%%%%%%%%%%%%%%%%%%%
\subsection{\label{subsec:flat_distr} Flat distributions}
%%%%%%%%%%%%%%%%%%%%%%%%%%%%%%%%%%%
%%%%%%%%%%%%%%%%%%%%%%%%%%%%%%%%%%%

\begin{figure}[tb!]
	\centering
	\includegraphics[width = 0.6\textwidth]{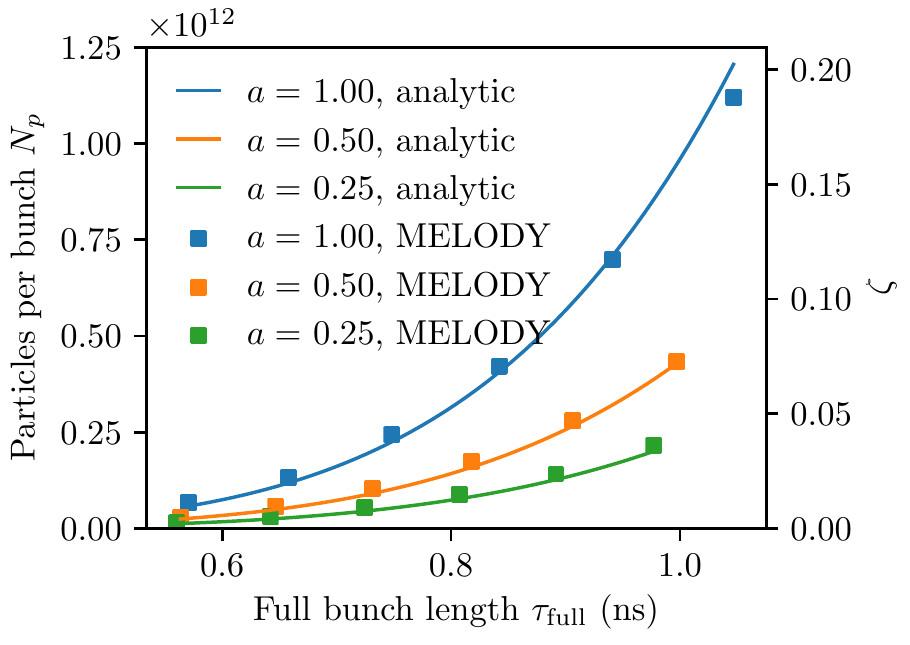}
	\caption{The LLD intensity threshold as a function of the full bunch length for the inductive impedance $\text{Im}Z/k= 0.07 \;\Omega$ ($\eta>0$) and the special particle distribution function~(\ref{eq:BI_distr_gen}) calculated for different parameters $a$ using MELODY (squares) and analytic expression~(\ref{eq:threshold_bi_approx}) (color lines) for $V_0 = 6$~MV.}
	\label{fig:comp_threshold_BI}
\end{figure}

Previous studies have pointed out that the LLD threshold is sensitive to the small-argument steepness of the distribution function~\cite{Burov2}.
Based on this fact, the dedicated measurements were performed at the Tevatron that allowed suppressing the bunch oscillations due to LLD~\cite{CYTanABurov2012}. Our studies show that the LLD threshold is indeed higher for the distribution function with $g^\prime(\E=0)=0$, used, for example, in Ref.~\cite{Balbekov-Ivanov1986}
\begin{equation}
    g(\E) = \left\{
  \begin{array}{ll}
     1-\frac{\E^2}{a\E^2_{\max}}, &  0 \leq \E < a\E_{\max} \\
    \frac{1}{1-a}\left(1- \frac{\E}{\E_{\max}}\right)^2, &  a\E_{\max} \leq \E < \E_{\max} \\
    0, & \text{elsewhere,}
  \end{array}\right.
  \label{eq:BI_distr_gen}
\end{equation}
where $0<a<1$.
For this case,
the diagonal matrix elements $G_{kk}$ (\ref{eq:Gkk}) are 
\begin{align}
    G_{kk} &=\frac{192i}{\pi(a^2-1) \phi^4_{\max}}\left[ J^2_0({}y) +J^2_1({}y)- \frac{J_0({}y)J_1({}y)}{{}y} \right. \nonumber\\
    &\left.-J^2_0\left({}y\sqrt{a} \right) -J^2_1\left({}y\sqrt{a} \right)+ \frac{J_0\left({}y\sqrt{a} \right)J_1\left({}y\sqrt{a} \right)}{{}y\sqrt{a}} \right].
    \label{eq:Gkk_bi}
\end{align}
For a very large argument ${}y\to\infty$, they simply become 
\begin{equation}
    \label{eq:gkk_bi_asympt}
    G_{kk}\approx \frac{384i}{\pi^2{}y\; \phi^4_{\max} (a+\sqrt{a})(a+1)},
\end{equation}
and approach zero as $1/y$. This is the main difference in comparison to the particle distributions of the binomial family, where the elements $G_{kk}$ saturate at some constant level. Calculating the sum of $G_{kk}$ to evaluate the LLD threshold using Eq.~(\ref{eq:LLD_threshold_general}), one gets a weak (logarithmic) dependence on the cutoff frequency. 
However, the semi-analytic calculations with MELODY show that the LLD threshold does not depend on the cutoff frequency, as can be seen in Fig.~\ref{fig:comp_threshold_BI}. 
The discrepancy can be understood after numerical evaluation of a small parameter $\varepsilon = \zeta/\phi^4_{\max}$, which, in this case,
is significantly larger than for the binomial distribution and thus expansion~(\ref{eq:matrix_det}) cannot be fully justified. Nevertheless, we propose the following analytic LLD threshold, based on the asymptotic behavior of $G_{kk}$ in Eq.~(\ref{eq:gkk_bi_asympt}),
\begin{equation}
    \label{eq:threshold_bi_approx}
    \zeta_\mathrm{th} \approx -\frac{\pi^2 (1+a)(\sqrt{a}+a) \phi^5_{\max} }{768}.
\end{equation}
It agrees well with the results from semi-analytic calculations (see Fig.~\ref{fig:comp_threshold_BI}). In particular, the LLD threshold for distribution (\ref{eq:BI_distr_gen}) with $a=1$ is by orders of magnitude higher than the one for the binomial distribution with $\mu = 2$ shown in Fig.~\ref{fig:comp_threshold_ind}.
We have also found that once the LLD threshold is reached, the response to a rigid-dipole perturbation can be stronger than for a bunch with the binomial distribution. 
Note, in operation this distribution was obtained as the result of the specific rf manipulations (e.g., rf phase modulation)~\cite{CYTanABurov2012,Shaposhnikova2014}.

%%%%%%%%%%%%%%%%%%%%%%%%%%%%%%%%%%%
%%%%%%%%%%%%%%%%%%%%%%%%%%%%%%%%%%%
\subsection{\label{subsec:inductance_below} Constant inductive impedance below transition energy}
%%%%%%%%%%%%%%%%%%%%%%%%%%%%%%%%%%%
%%%%%%%%%%%%%%%%%%%%%%%%%%%%%%%%%%%

\begin{figure}[tb!]
	\centering
	\includegraphics[width = 0.6\textwidth]{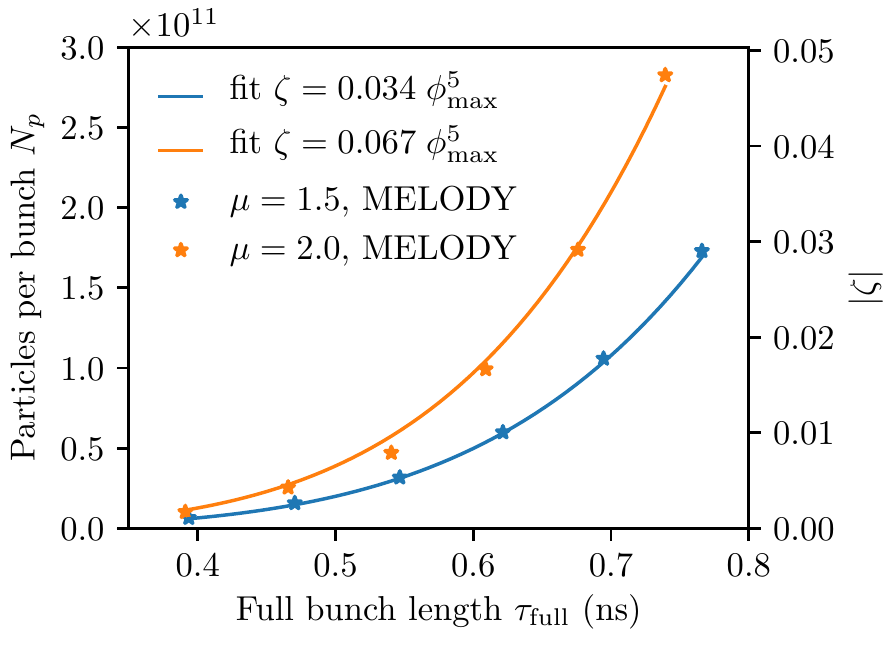}
	\caption{The LLD intensity threshold as a function of the full bunch length for the inductive impedance $\text{Im}Z/k= 0.07 \;\Omega$ below transition ($\eta<0$), calculated for the binomial distribution with different values of $\mu$ using MELODY (squares) and fitted functions (color lines) for $V_0 = 6$~MV.}
	\label{fig:comp_threshold_neg_imzn}
\end{figure}

Here, we will show that the LLD threshold does not depend on the cutoff frequency for the case of the inductive impedance below transition or the space charge above transition (i.e. $\eta \;\text{Im}Z/k<0$).
In this case, at the threshold, the van Kampen mode emerges below the minimum incoherent frequency, and we can calculate elements $G_{kk}$ for $\Omega = \omega_{s0}(1-\E_{\max}/8)$,
\begin{align}
    G_{kk} &= -\frac{32i\mu(\mu+1)}{\pi \phi^4_{\max}}\int^1_0 xdx \frac{\left(1-x^2\right)^{\mu-1}}{1-x^2} J^2_1\left(\frac{kx}{h}\sqrt{2\E_{\max}} \right) \nonumber \\
    &= i\frac{4(\mu+1)}{\pi(\mu-1) \phi^4_{\max}} {}y^2 {}_1 F_2 \left(\frac{3}{2}; 3, \mu+1 ; -{}y^2 \right),\label{eq:Gkkneg}
\end{align}
for $\mu>1$.
For a very large argument ${}y\to\infty$, they can be expressed as
\begin{equation}
    \label{eq:Gkk_neg_asympt}
    G_{kk} \approx -\frac{16i\mu(\mu+1) \Gamma(\mu-1)}{\pi^{3/2} y\;\phi^4_{\max} \Gamma(\mu-1/2)},
\end{equation}
and also approach zero, similar to the case of flat bunches. Thus we expect that the LLD threshold does not depend on the cutoff frequency. Indeed, this can be seen in Fig.~\ref{fig:comp_threshold_neg_imzn}, where the semi-analytic calculations using MELODY are shown for different values of $\mu$.
There is a strong dependence on the tails of distribution and for the moment only fitted thresholds are proposed:  $|\zeta_\mathrm{th}| \approx 0.034 \phi^5_{\max}$ for $\mu=1.5$, and  $|\zeta_\mathrm{th}| \approx 0.067 \phi^5_{\max}$ for $\mu = 2$.

%%%%%%%%%%%%%%%%%%%%%%%%%%%%%%%%%%%%%%%%%%%%%%%%%%%%%%%%%%%%
\section{\label{sec:conclusions}Conclusions}
%%%%%%%%%%%%%%%%%%%%%%%%%%%%%%%%%%%%%%%%%%%%%%%%%%%%%%%%%%%%
Loss of Landau damping (LLD) in the longitudinal plane can be an important performance limitation of the existing and future storage rings. 
In the present paper, the criterion of the emerged van Kampen mode  
was used to determine the LLD thresholds for different beam and machine parameters. In particular, we were able to derive a general analytic expression for the LLD threshold of the dipole oscillations in a single rf system exploiting the Lebedev equation.

Contrary to the previous studies, we have found that for a particle distribution of the binomial family, a constant inductive impedance Im$Z/k$ above transition (the LHC case) or capacitive (space charge) below leads to a zero LLD threshold. In fact, it becomes inversely proportional to the cutoff frequency $f_c$ 
for $f_c\gg 1/\tau_\mathrm{full}$ ($\tau_\mathrm{full}$ is the full bunch length). We have confirmed this dependence by solving the Lebedev matrix equation semi-analytically as well as using Oide-Yokoya method, also showing that both numerical methods agree extremely well. 
The finite LLD threshold obtained in the previous studies using macro-particle simulations or the Oide-Yokoya method is due to some finite maximum frequency of inductive impedance naturally existing in numerical calculations.
In reality, there is always the cutoff frequency of the beam pipe limiting constant reactive impedance $\text{Im}Z/k$ at some value, with the proper decay law at higher frequencies. 
Classical dependence of the LLD threshold on the bunch length in the fifth power is justified only in the low cutoff-frequency limit ($f_c\le 1/\tau_\mathrm{full}$). The LLD threshold obtained by Sacherer and Hofmann-Pedersen approaches can be reproduced when $f_c \approx  1/\tau_\mathrm{full}$.  This dependence changes to the power of four for the case of the higher cutoff frequency ($f_c \gg 1/\tau_\mathrm{full}$).
Introducing a new definition of effective impedance,  we have found for a broad-band resonator model similar dependency on the resonant frequency and the bunch length that were also confirmed by semi-analytic calculations using code MELODY 
as well as macro-particle simulations with code BLonD.

Another important step in the understanding of LLD is the evaluation of the beam response to the rigid-dipole perturbation (a kick) which may lead in accelerator to undamped bunch oscillations. To reconstruct the bunch offset evolution,
 this perturbation was expanded on the basis of van Kampen modes and then tracked in the time domain. 
We have shown for the binomial distribution that
even though the LLD threshold is lower for  higher cutoff/resonant frequencies, an impact on the beam at the LLD threshold of impedance with higher frequencies is smaller,
also with a strong dependence on the impedance model.
On the other hand, for intensities well above the LLD threshold, the beam response is mostly defined by the bunch parameters. Again, these results were confirmed by simulations that required to use tens of millions of macro-particles and sufficient slicing in the induced voltage calculations to cope with the numerical noise. We also have found that the amplitude of bunch oscillations after a phase offset has a strong dependence on the bunch length (in the sixth power). Finally, our calculations are in good agreement with available measurements in the LHC. 

There is also a special class of particle distribution functions, with zero derivative in the center of the bunch, which leads to a high and  finite LLD threshold for a constant inductive impedance Im$Z/k$ above transition energy (and capacitive below).
Moreover, we also shortly discussed the case of a constant inductive impedance Im$Z/k$ below transition energy (space impedance above) that results in the finite LLD threshold for the binomial distributions.
A similar analysis can be also applied for higher-order azimuthal modes and different forms of the bare rf potential (double rf system). This could be a subject of further studies.
\noindent

\section*{ACKNOWLEDGEMENTS}
We thank Alexey Burov for useful discussions and comments.

\appendix
\begin{appendices}
%%%%%%%%%%%%%%%%%%%%%%%%%%%%%%%%%%%%%%%%%%%%%%%%%%%
\section{\label{annex:A} Harmonics of the induced voltage}
%%%%%%%%%%%%%%%%%%%%%%%%%%%%%%%%%%%%%%%%%%%%%%%%%%%
The following definition of impedance is used in the present work~\cite{Zotter}
\begin{equation}
    \label{eq:impdance}
    Z(\omega) = \int_{-\infty}^\infty \frac{d\phi}{\omega_\mathrm{rf}} \, W(\phi) e^{-i \omega \phi/\omega_\mathrm{rf}},
\end{equation}
with corresponding wake function $W(\phi)$
\begin{equation}
    \label{eq:wake_function}
    W(\phi) = \int_{-\infty}^\infty \frac{d\omega }{2\pi} \, Z(\omega) e^{i \omega \phi/\omega_\mathrm{rf}}.
\end{equation}
We use the convention $W(\phi) = 0 $ for $\phi < 0$ due to causality.

Below we will derive the induced voltage for the stationary and the time-dependent cases.
The stationary part of the induced voltage for a single bunch with intensity $N_p$ is related to the normalized line density $\lambda(\phi)$ as~\cite{Chao}
\begin{equation}
    \label{eq:induced_voltage_wake_st}
    V_\mathrm{ind}(\phi) = -q N_p  \int_{-\infty}^\infty d\varphi \; \lambda(\varphi) \sum_{k=0}^\infty W(\phi - \varphi +  2\pi h k),
\end{equation}
where the normalization  $ \int_{-\pi h}^{\pi h}d\phi \lambda(\phi) = 1$ was imposed.
% , while stationary situation can be simply obtained from the results by setting coherent mode frequency $\Omega$ to zero.
Because of causality, sum can also be extended for $k<0$. Inserting the relation between wake function and impedance (\ref{eq:wake_function}) one gets
\begin{equation}
    \label{eq:induced_voltage_wake_st2}
    V_\mathrm{ind}(\phi) = -q N_p  \int_{-\infty}^\infty d\varphi \, \lambda(\varphi) \int_{-\infty}^\infty \frac{d\omega}{2\pi} Z(\omega) e^{i\omega(\phi-\varphi)/\omega_\mathrm{rf}} \sum_{k=-\infty}^\infty e^{i 2\pi k \omega/\omega_0}. 
\end{equation}
Using here the Dirac comb relation
\begin{equation}
    \label{eq:dirac_st}
    \sum_{k=-\infty}^\infty e^{i 2\pi k \omega /\omega_0} = \omega_0 \sum_{k=-\infty}^\infty \delta\left[\omega - k\omega_0 \right],
\end{equation}
we can present the induced voltage as
\begin{equation}
    \label{eq:induced_voltage_wake_st3}
    V_\mathrm{ind}(\phi) = -q N_p h \omega_0  \sum_{k=-\infty}^\infty Z_k \lambda_k e^{i k\phi/h} = \sum_{k=-\infty}^\infty V_k e^{i k\phi/h},
\end{equation}
where $Z_k = Z(k \omega_0)$ and the harmonic of the line density
\begin{equation}
    \label{eq:harmonic_line_density_time_st}
    \lambda_k= \frac{1}{2\pi h} \int_{-\pi h}^{\pi h} d\phi \, \lambda(\phi) e^{-i k\phi/h}.
\end{equation}

Assuming the perturbation of the line density in the form $\Tilde{\lambda}(\phi,\Omega,t) = \Tilde{\lambda}(\phi,\Omega) e^{-i\Omega t}$,
% $\Tilde{\F} (\E,\psi,t) = \Tilde{\F} (\E,\psi,\Omega) e^{-i\Omega t}$, the corresponding line density is
% \[
% \Tilde{\lambda}(\phi,\Omega,t) = \Tilde{\lambda}(\phi,\Omega) e^{-i\Omega t} = \int_{-\infty}^\infty d\dot{\phi} \Tilde{\F}(\E,\psi,\Omega) e^{-i\Omega t}.
% \]
the induced voltage can be expressed using the convolution of line density with wake function, similarly to Eq.~(\ref{eq:induced_voltage_wake_st}),~\cite{Chao}
\begin{equation}
    \label{eq:induced_voltage_wake1}
    \Tilde{V}_\mathrm{ind}(\phi,t) = -q N_p  \int_{-\infty}^\infty d\varphi \, \Tilde{\lambda}(\varphi,\Omega) \sum_{k=0}^\infty W(\phi - \varphi +  2\pi h k) e^{-i \Omega (t + k T_0)}.
\end{equation}
Again, extending the sum for $k<0$ due to causality and inserting the definition of the wake function (\ref{eq:wake_function}), we obtain
\begin{equation}
    \label{eq:induced_voltage_wake2}
    \Tilde{V}_\mathrm{ind}(\phi,t) = -q N_p e^{-i \Omega t}  \int_{-\infty}^\infty d\varphi \Tilde{\lambda}(\varphi,\Omega) \int_{-\infty}^\infty \frac{d\omega}{2\pi} Z(\omega) e^{i\omega(\phi-\varphi)/\omega_\mathrm{rf}} \sum_{k=-\infty}^\infty e^{i 2\pi k (\omega- \Omega)/\omega_0} .
\end{equation}
In this case, the Dirac comb relation is
\begin{equation}
    \label{eq:dirac}
    \sum_{k=-\infty}^\infty e^{i 2\pi k \left(\omega - \Omega\right)/\omega_0} = \omega_0 \sum_{k=-\infty}^\infty \delta\left[\omega - \left(k\omega_0 + \Omega \right) \right].
\end{equation}

Finally, we get the induced voltage in the following form
\begin{equation}
    \label{eq:induced_voltage_wake3}
    \Tilde{V}_\mathrm{ind}(\phi,t) = -q N_p h \omega_0 e^{-i \Omega t} \sum_{k=-\infty}^\infty Z_k(\Omega) \Tilde{\lambda}_k(\Omega) e^{i k\phi/h} e^{i \Omega \phi /\omega_\mathrm{rf}} = e^{-i \Omega t}\sum_{k=-\infty}^\infty \Tilde{V}_k(\Omega) e^{i k\phi/h} e^{i \Omega \phi /\omega_\mathrm{rf}},
\end{equation}
where $Z_k(\Omega) = Z(k \omega_0 + \Omega)$ and the harmonic of the line density perturbations
\begin{equation}
    \label{eq:harmonic_line_density_time}
    \Tilde{\lambda}_k(\Omega) = \frac{1}{2\pi h} \int_{-\pi h}^{\pi h} d\phi \, \Tilde{\lambda}(\phi, \Omega) e^{-i k\phi/h} e^{-i\Omega\phi/\omega_\mathrm{rf}}.
\end{equation}
In the present work we consider the case when $\Omega \approx m \omega_{s0} \ll \omega_0$ for relatively low azimuthal modes  $m = 1,2,3, ... $, so the phase factor $e^{\pm i \Omega \phi/ \omega_\mathrm{rf}}$ in Eqs.~(\ref{eq:induced_voltage_wake3}, \ref{eq:harmonic_line_density_time}) is neglected for the rest of derivations.

%%%%%%%%%%%%%%%%%%%%%%%%%%%%%%%%%%%%%%%%%%%%%%%%%%%%%%%%%%%%%%%%%%%%%
\section{\label{annex:B} Stationary potential}
%%%%%%%%%%%%%%%%%%%%%%%%%%%%%%%%%%%%%%%%%%%%%%%%%%%%%%%%%%%%%%%%%%%%%
The iterative procedure~\cite{burov2012van} can be used to obtain the solution for a stationary situation with intensity effects included (also called potential well distortion).  In a single rf system, the total potential can explicitly be written using Eqs.~(\ref{eq:induced_voltage},\ref{eq:potential})
\begin{flalign}
    U_t(\phi) = U_\mathrm{rf}(\phi) + U_\mathrm{ind}(\phi) 
    &= \frac{\cos(\phi_{s0} + \Delta \phi_s)- \cos(\phi_{s0} + \phi)}{\cos\phi_{s0}} - (\phi-\Delta \phi_s)\tan\phi_{s0} \nonumber \\
    &+ \frac{i q N_p \,h^2 \, \omega_0 }{V_0 \cos\phi_{s0}}\sum_{k=-\infty}^{\infty}\frac{Z_k}{k} \lambda_k \left( e^{i \frac{k}{h} \phi} - e^{i \frac{k}{h} \Delta \phi_s} \right),  \label{eq:total_potential}
\end{flalign}
and it depends on  harmonics of the line density $\lambda_k$.
% In turn, the  normalized line density depends on the total potential as
% The general stationary particle distribution can be an arbitrary function of $\E$.
% The potential and line density can be found from an iterative process, where

As the first step, we calculate the potential $U_{t,0}$ from Eq.~(\ref{eq:total_potential}) and the line density $\lambda_0$  from either Eq.~(\ref{eq:norm_line_density}) or Eq.~(\ref{eq:line_density_binom}) for $N_p=0$ and $\Delta \phi_s = 0$ (no intensity effects). Then, at given step $n$:
\begin{align}
    \label{eq:iterations}
    U_{t,n} &=(1-\epsilon) U_{t,n-1} + \epsilon U_t(\lambda_{n-1}); \nonumber \\
    U_{t,n} &= U_{t,n} - \min(U_{t,n}); \\
    \lambda_n &= \lambda(U_{t,n}); \text{  } n = 1,2,... \nonumber
\end{align}
The solution of this system of equations,
if it exists, can be found for a sufficiently small convergence parameter $\epsilon>0$. 

In the case of the constant inductive impedance $Z_k = ik \text{Im}Z/k$, Eq.~(\ref{eq:total_potential}) can be simplified for the binomial distribution by using Eq.~(\ref{eq:line_density_binom}),
to obtain an implicit form of the total potential
\begin{equation}
    \label{eq:total_potential_ind}
    U_t(\phi) = U_\mathrm{rf}(\phi) - \zeta \left[\lambda(\phi) -\lambda(0) \right] = U_\mathrm{rf}(\phi) - \zeta \lambda(0) \left\{ \left[1 - U_t(\phi)/\E_{\max}\right]^{\mu + 1/2} - 1\right\},
\end{equation}
Here we used the dimensionless parameter 
\begin{equation}
    \label{eq:xi_annex}
     \zeta = -\frac{ q N_p \, h^2 \, \omega_0 \text{Im} Z/n }{V_0 \cos\phi_{s0}}.
\end{equation}
The total potential can be obtained analytically for $\mu = 1/2$ as in  Hofmann-Pedersen approach~\cite{Hofmann1979}, but also for $\mu =$ 0, 1 and 3/2, for example, where either quadratic or cubic equation needs to be solved. In general, the first derivative can easily be found as
\begin{equation}
    \label{eq:derivative potential}
    \frac{d U_t(\phi)}{d\phi} = \frac{1}{ B (\phi)} \frac{dU_\mathrm{rf}(\phi)}{d \phi},
\end{equation}
where 
%the function in denominator
\begin{equation}
    \label{eq:denominator}
    B (\phi) = 1 - \zeta \frac{\lambda(0) (\mu+1/2)}{\E_{\max}}\left[1- \frac{U_t(\phi)}{\E_{\max}}\right]^{\mu - 1/2}.
\end{equation}
The second derivative that defines the small-amplitude synchrotron frequency is
%can be found similarly
\begin{equation}
    \label{eq:second_derivative potential}
    \frac{d^2 U(\phi)}{d\phi^2} = \frac{1}{ B (\phi)} \frac{d^2U_\mathrm{rf}(\phi)}{d \phi^2} -\zeta\frac{\lambda(0) (\mu^2 -1/4)}{[\E_{\max} B (\phi)]^2} \frac{dU_\mathrm{rf}(\phi)}{d \phi}\frac{dU_t(\phi)}{d \phi}.
\end{equation}
All higher-order derivatives can be calculated recursively.

The case of $\mu = 0$ is of particular interest. From the above equations one can see that all derivatives vanish at the maximum and minimum particle excursions in the bunch, $\phi_{\min}$ and $\phi_{\max}$, respectively. In means that the synchrotron period (\ref{eq:synchrotron_freq}) is infinite for that trajectory and thus $\omega_s(\E_{\max}) = 0$.

\end{appendices}
\nocite{apsrev41Control}
\bibliographystyle{apsrev4-1}
\bibliography{Landau_damping.bib}

\end{document}